\pdfoutput=1
\documentclass[aps,pra,amsmath,amssymb,floatfix,nofootinbib,notitlepage, superscriptaddress,10pt]{revtex4-1}
\usepackage{amsthm}
\usepackage{latexsym}
\usepackage{amsfonts}
\usepackage{bbm,dsfont}
\usepackage{graphicx}
\usepackage[usenames,dvipsnames]{xcolor}
\usepackage{amsthm}
\usepackage{latexsym}
\usepackage{amsfonts}
\usepackage{bbm,dsfont}
\usepackage{graphicx,xcolor}
\usepackage{braket}
\usepackage{hyperref}
\definecolor{bottle_green}{RGB}{0,106,78}
\definecolor{celadon_green}{RGB}{47,132,124}
\definecolor{emerald}{RGB}{80,220,100}
\definecolor{jade}{RGB}{0,168,107}
\hypersetup{colorlinks,
linkcolor=bottle_green,
citecolor=celadon_green,
urlcolor=emerald,
anchorcolor=jade}
\newcommand{\R}{\mathbb{R}}
\newcommand{\C}{\mathbb{C}}
\newcommand{\N}{\mathbb{N}}
\newcommand{\Z}{\mathbb{Z}}
\newcommand{\BH}{\mathcal{B}}
\newcommand{\hil}{\mathcal{H}}
\newcommand{\Id}{\mathbb{I}}
\newcommand{\tr}[1]{\mathrm{Tr}\left[ {#1} \right]} 
\newcommand{\Tr}[2]{\mathrm{Tr}_{#1}\left[ {#2} \right]} 
\newcommand{\p}{\mathrm{p}}
\newcommand{\q}{\mathrm{q}}
\newcommand{\T}{\mathrm{T}}
\newcommand{\rmd}{\mathrm{d}}

\newtheorem{definition}{Definition}
\newtheorem{technical_result}{Technical Result}

\begin{document}
\title{Energetic instability of passive states in thermodynamics}
\author{Carlo Sparaciari}\email{carlo.sparaciari.14@ucl.ac.uk}
\affiliation{Department of Physics and Astronomy, University College London,
London WC1E 6BT, United Kingdom}
\author{David Jennings}\email{david.jennings@physics.ox.ac.uk}
\affiliation{Department of Physics, University of Oxford, Oxford, OX1 3PU, United Kingdom}
\affiliation{Department of Physics, Imperial College London, London SW7 2AZ, United Kingdom}
\author{Jonathan Oppenheim}\email{j.oppenheim@ucl.ac.uk}
\affiliation{Department of Physics and Astronomy, University College London,
London WC1E 6BT, United Kingdom}
\begin{abstract}
Passivity is a fundamental concept in thermodynamics that demands a quantum system's energy cannot
be lowered by any reversible, unitary process acting on the system. In the limit of many such systems, passivity
leads in turn to the concept of complete passivity, thermal states, and the emergence of a thermodynamic
temperature. In contrast, here we need only consider a single system and show that every
passive state except the thermal state is unstable under a weaker form of reversibility. More precisely, we show that given a single
copy of any athermal quantum state we may extract a maximal amount of energy from the state when we
can use a machine that operates in a reversible cycle and whose state is left unchanged. This means that
for individual systems the only form of passivity that is stable under general reversible processes is complete
passivity, and thus provides a single-shot and more physically motivated identification of thermal states and the emergence of temperature.
The machine which extracts work from passive states exploits the fact that one can find a subspace which acts as a virtual hot reservoir,
  and a subspace which acts as a virtual cold reservoir. We show that an optimal amount of work can be extracted, and that the machine
  operates at the Carnot efficiency between pairs of virtual reservoirs.
\end{abstract}
\date{\today}
\maketitle
\section{Introduction}
\label{into_passive_state}
Within thermodynamics, heat engines are devices that operate in a thermal context so as to extract
ordered energy in the form of work. The canonical scenario involves an engine that operates cyclically
between two temperatures $T_{\text{hot}}$, $T_{\text{cold}}$ and performs a quantity of mechanical work.
To do so the engine absorbs heat from the hot reservoir, converts some of this energy to mechanical work
and releases heat into the cold reservoir in accordance with the Second Law of thermodynamics. The
largest possible efficiency, $\eta = 1 - \frac{T_{\text{cold}}}{T_{\text{hot}}}$, occurs for the reversible
Carnot engine~\cite{carnot_reflections_1824, callen_thermodynamics_1985} and provides a fundamental
thermodynamic bound on the amount of ordered energy that can be obtained. Carnot engines, and more
in general heat engines, have been extensively studied in the microscopic regime~\cite{scully_extracting_2003, scovil_three-level_1959, geusic_quantum_1967, alicki_quantum_1979, howard_molecular_1997, geva_classical_1992, hanggi_artificial_2009, feldmann_quantum_2006, rousselet_directional_1994, faucheux_optical_1995, linden_how_2010, horodecki_fundamental_2013, brunner_virtual_2012, tajima_optimal_2014, ito_optimal_2016, verley_unlikely_2014,
gardas_thermodynamic_2015, uzdin_collective_2015, uzdin_quantum_2016, kosloff_quantum_2016,
woods_maximum_2015, frenzel_quasi-autonomous_2016, lekscha_quantum_2016, niedenzu_operation_2016}
(as well as, of course, in the macroscopic regime).
\par
However, the issue of ordered energy extraction can also be considered in scenarios in which no notion of temperature exists, and can provide a broader notion of equilibrium states. For example, more general equilbrium states can occur in physical realisations when a system has been perturbed and has not had enough time to fully thermalise. They can also arise in the context of non-equilibrium steady states \cite{seifert2012stochastic}. Given a quantum system in a state $\rho$ one can ask if it is possible to extract energy from it  solely by performing a reversible unitary transformation on the system. The largest amount of ordered energy that can be extracted (the ``ergotropy", see Refs.~\cite{allahverdyan_maximal_2004, alicki_entanglement_2013, sparaciari_resource_2016}) depends non-trivially on the quantum state. If no energy can be extracted in this way then $\rho$ is called \emph{passive}~\cite{pusz_passive_1978, lenard_thermodynamical_1978, perarnau-llobet_extractable_2015, perarnau-llobet_most_2015, hovhannisyan_entanglement_2013} and constitutes a primitive form of equilibrium.
\par
In this work, we consider a scenario that is intermediate between the above two contexts, and is motivated
by the fact that a work extraction machine should be considered as a system which is involved in the process.
Our core question is whether there exist passive states $\rho_S$ for which energy can be
extracted if one performs a reversible unitary process over the system $S$ together with a
second quantum system $M$, which starts and finishes in the same quantum state $\rho_M$.
This second quantum system is the machine, such as the working body in a Carnot cycle, which
undergoes a cyclic evolution. This class of processes is reminiscent of the ones taking place inside heat engines, and
has been termed a {\it catalytic thermal operation}~\cite{brandao_second_2013}.
\par
Indeed, the system $S$ described by a passive state represents \emph{both} the reservoirs,
while the additional system $M$ is the machine which exchanges energy with the reservoirs
in a cyclic manner. Due to the fact that microscopic heat engines can nowadays be realised
in the laboratory~\cite{scovil_three-level_1959, rousselet_directional_1994, faucheux_optical_1995, howard_molecular_1997, schulman_molecular_1999, scully_quantum_2002},
it seems reasonable to extend the analysis on passive states to the case
in which this broader class of operations is allowed. Crucially, this analysis has fundamental implications
for the notion of passivity. In fact, if energy can be extracted from a passive state with these
reversible processes, and no entropy is generated, then it seems that associating passivity of
the state is a restricted idealisation, unstable under this simple extension.
\par
The paper is structured as follows. In Sec.~\ref{passive_properties} we provide the definition of passive and
  completely passive states, together with their description in terms of virtual temperatures. Any three levels of a passive state can be thought of as containing
  a subsystem which acts as a virtual hot reservoir, and one which acts as a virtual cold one.
Sec.~\ref{general_cycle}
provides the description of a protocol involving finite-sized engines which enable us to extract work in a cyclic way from a single
passive state. In Sec.~\ref{activation_region} we prove that in fact any passive state which is not thermal can be activated by such an engine. Finally in Sec.~\ref{asympt_machine}
we show that such an activation can be done optimally via a quasi-static, reversible process with zero generation of entropy. We provide a simple expression for the amount of work which can be extracted from an arbitrary passive state, Eq.~\eqref{eq:optimalwork}, and show that our machine can operate at the efficiency of a Carnot engine operating
  between two virtual heat reservours. We conclude that the only states in the single-shot regime that are stable under general reversible processes are thermal states.
\section{Passive states}
\label{passive_properties}
Consider a finite-dimensional quantum system associated with the Hilbert space $\hil \equiv \C^d$
(a \emph{qudit}), with Hamiltonian $H = \sum_{i=0}^{d-1} E_i \ket{i}\bra{i}$, and described by the
state $\rho$. We say that the state $\rho$ is \emph{passive} iff its average energy cannot be lowered
by acting on it with unitary operations, that is,
\begin{equation} \label{passive_condition}
\tr{H \, \rho} \leq \tr{H \, U \rho \, U^{\dagger}} \ , \ \forall \, U \in \BH(\hil) \ , \ U U^{\dagger} = U^{\dagger} U = \Id.
\end{equation}
This implies that no work can be extracted from the state via a unitary process, since by conservation of energy, lowering
  the energy of a system would mean that this energy has been transfered to a work storage device.
\par
We can also introduce a more restrictive notion of passivity. Let us consider $n \in \N$ independent and identically
distributed (i.i.d.) copies of our system, with a total Hamiltonian $H^{(n)} = \sum_{i=1}^{d} H_i$, where each
$H_i$ is a single-system Hamiltonian acting on a different copy of the system. The state of this global system
is described by $\rho^{\otimes n}$. Then, we say that the state $\rho$ is \emph{completely passive} if
and only if the state $\rho^{\otimes n}$ is passive for all $n \in \N$. It can be shown \cite{pusz_passive_1978}
that the completely passive states of a system with Hamiltonian $H$ are the ones satisfying the KMS
condition~\cite{kubo_statistical-mechanical_1957, martin_theory_1959, haag_equilibrium_1967}. Specifically,
these states are the ground state and the thermal states with temperature $\beta \geq 0$, that is,
$\tau_{\beta} = e^{- \beta H}/Z$ with $Z = \tr{e^{- \beta H}}$. {\it Any state which is not of this form, is called {\it athermal}}.
\begin{figure}[!ht]
\center
\includegraphics[width=0.8\textwidth]{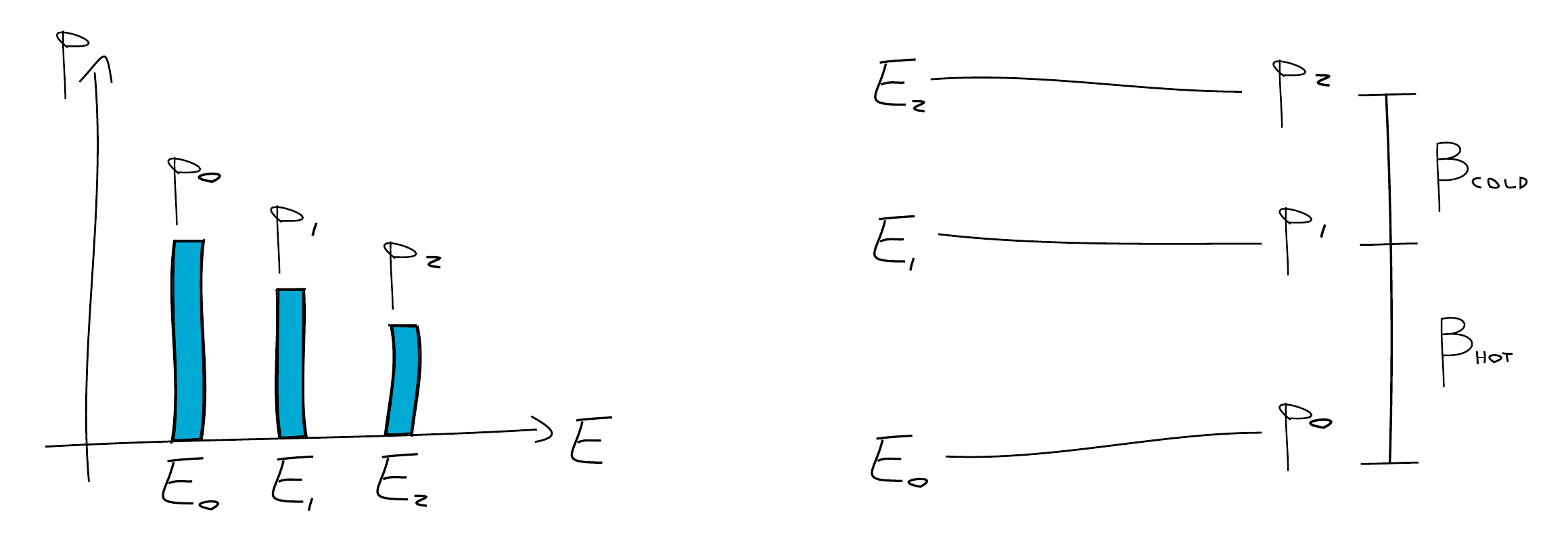}
\caption{(Left) The spectrum of a qutrit passive state $\rho = \sum _{i=0}^2 p_i
\ket{i}\bra{i}$ over the eigenbasis of its Hamiltonian $H = \sum _{i=0}^2 E_i
\ket{i}\bra{i}$. The occupation probabilities are ordered in a decreasing order,
from the one associated with the ground state of $H$ to the one associated with
the maximally excited one, as described in Eq.~\eqref{passive_stair}.
(Right) A passive state can equally be described by {\it virtual temperatures}. Indeed, we
can associate a virtual temperature to each pair of eigenstates of $\rho$, as shown
in Eq.~\eqref{virtual_temperature}. In the plot, the pair of levels $\ket{0}$ and
$\ket{1}$ is associated with the hot temperature $\beta_{\text{hot}}^{-1}$, while
the pair  $\ket{1}$ and $\ket{2}$ is associated with the cold temperature
$\beta_{\text{cold}}^{-1}$.}
\label{fig2:passive_state}
\end{figure}
\par
A characterization of all passive states can be easily obtained. A system in a passive state is such that the ground state has the highest probability of being occupied, and
the probability of occupation decreases as the energy associated with the eigenstate of $H$ increases,
Fig.~\ref{fig2:passive_state}, left plot. Specifically, a state $\rho$ is passive iff $\rho= f(H)$, where $f$ is a monotone non-increasing function. Simply put, this means that the state can expressed as
\begin{equation}
\label{passive_stair}
\rho = \sum_{i=0}^{d-1} \p_i \ket{i}\bra{i} \ , \ \text{such that} \ \p_i \geq \p_{i+1} \ \forall \, i = 0, \ldots , d-2,
\end{equation}
where $\left\{ \ket{i} \right\}_{i=0}^{d-1}$ are the eigenvectors of $H$, ordered so that $E_i \leq E_{i+1}$ for all $i$ (for the case of equal energies $E_i=E_{i+1}$ we must make an additional stability assumption to ensure that $p_i = p_{i+1}$).
\par
We can describe the probability distribution of the passive state $\rho$ by using \emph{virtual}
temperatures~\cite{brunner_virtual_2012, skrzypczyk_passivity_2015}. In fact, for any
given passive state, we can associate a (non-negative) virtual temperature with each
pair of its eigenstates. For example, if we consider the pair $\left( \ket{i}, \ket{j} \right)$,
we define the virtual temperature associated with them as the $\beta_{ij}^{-1} \geq 0$
such that
\begin{equation}
\label{virtual_temperature}
\frac{\p_i}{\p_j} =: e^{- \beta_{ij} \left( E_i - E_j \right)},
\end{equation}
where $\p_i$ is the probability of occupation of the state $\ket{i}$, and $E_i$ is the energy
associated with the state (similarly for $j$). Thus, each pair of states can be regarded as an effective thermal
state at a specific temperature. When all pairs of states has the same virtual temperature, we
have that the passive state is completely passive, that is, it is the thermal state of $H$ at that
temperature.
\section{The core protocol}
\label{general_cycle}
We now introduce an engine that extracts work by acting individually on a passive state. The engine is
composed by two main elements, namely, a qudit ``machine" system with trivial Hamiltonian ($H=0$),
and a particular passive state. It suffices to consider qutrit systems, as a similar construction works more
generally. The qutrit is assumed to have a Hamiltonian
\begin{equation} \label{hamiltonian}
H_P = \sum_{i=0}^2 E_i \ket{i}\bra{i}_P,
\end{equation}
and $E_i \leq E_{i+1}$. The qutrit system is described by the state
\begin{equation} \label{qutrit_passive_state}
\rho_P = \sum_{i=0}^2 \p_i \ket{i}\bra{i}_P,
\end{equation}
where $\p_i \geq \p_{i+1}$ (Fig.~\ref{fig2:passive_state}, left plot). 
\par
In the following we assume that the passive state $\rho_P$ is described by the virtual temperature
$T_{\text{hot}} = \beta_{\text{hot}}^{-1} > 0$ associated with the pair of eigenstates $\left( \ket{0}_P ,
\ket{1}_P \right)$, and the virtual temperature $T_{\text{cold}} = \beta_{\text{cold}}^{-1} > 0$ associated with
the pair $\left( \ket{1}_P , \ket{2}_P \right)$. We assume for simplicity that $T_{\text{hot}} > T_{\text{cold}}$, but a similar analysis applies for $T_{\text{hot}} < T_{\text{cold}}$. In the Supplemental Material, Sec.~I, the cycle is presented in full detail. The relation between
the probability distribution of $\rho_P$ and the temperatures $T_{\text{hot}}$ and $T_{\text{cold}}$
is given by
\begin{subequations} \label{hot_cold_temperatures}
\begin{align}
\frac{\p_1}{\p_0} &=: e^{- \beta_{\text{hot}} \Delta E_{10}}, \\
\frac{\p_2}{\p_1} &=: e^{- \beta_{\text{cold}} \Delta E_{21}},
\end{align}
\end{subequations}
where $\Delta E_{10} = E_1 - E_0 \geq 0 $, and $\Delta E_{21} = E_2 - E_1 \geq 0$.
Thus, the pair of states $\left( \ket{0}_P , \ket{1}_P \right)$ can be viewed as representing a ``hot virtual reservoir",
while the pair of states $\left( \ket{1}_P , \ket{2}_P \right)$ represent a ``cold virtual reservoir". It is worth noting
that the other pair of states, $\left( \ket{0}_P , \ket{2}_P \right)$, is associated with a virtual temperature
that is intermediate between $T_{\text{cold}}$ and $T_{\text{hot}}$, as we can easily verify from
Eqs.~\eqref{hot_cold_temperatures}.
\par
\begin{figure}[!ht]
\centering
\includegraphics[width=0.8\textwidth]{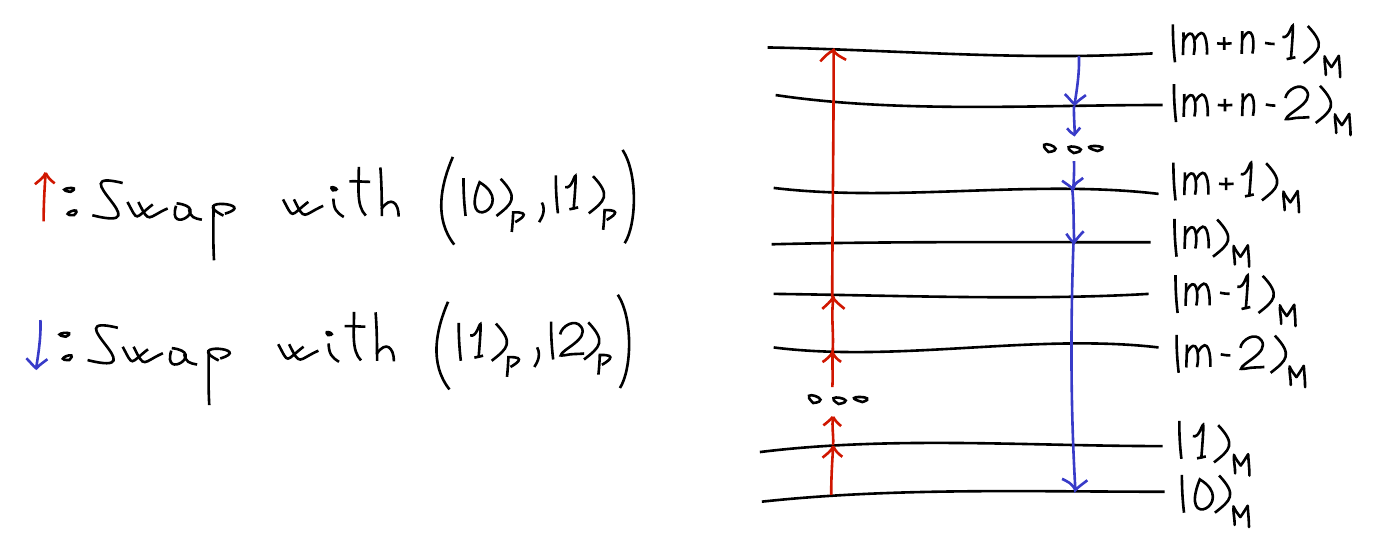}
\caption{The cycle is represented in a pictorial way over the eigenstates of the $d$-dimensional machine (where $d = m + n$). Notice that the machine has a trivial Hamiltonian,
and we order the eigenstates only to simplify the visualisation of the cycle. The upward arrow
connecting two eigenstates of the machine represents a swap between these two states
and the pair $\left( \ket{0}_P , \ket{1}_P \right)$ of the passive state. The downward arrow connecting
two eigenstates of the machine represents a swap between this pair and the pair
$\left( \ket{1}_P , \ket{2}_P \right)$ of the passive state. We initially perform $m-1$ swaps between
$\left( \ket{0}_P , \ket{1}_P \right)$ and $\left\{\left( \ket{j}_M , \ket{j+1}_M \right) \right\}_{j=0}^{m-2}$,
and one swap between $\left( \ket{0}_P, \ket{1}_P \right)$ and $\left( \ket{m-1}_M , \ket{m+n-1}_M \right)$.
Then, we perform $n-1$ swaps between $\left( \ket{1}_P , \ket{2}_P \right)$ and $\left\{ \left( \ket{j}_M ,
\ket{j+1}_M \right) \right\}_{j=m}^{m+n-2}$, and one swap between $\left( \ket{1}_P , \ket{2}_P \right)$
and $\left( \ket{0}_M , \ket{m}_M \right)$. If we consider the arrow representation of swaps,
we can see that the cycle is close, and this allows us to recover the local state of the
machine $M$ while extracting work.}
\label{fig3:catalyst_diag}
\end{figure}

The engine extracts work by means of the following cycle. A single system, described by
the state $\rho_P$, is put in contact with the
machine, described by the state $\rho_M = \sum_{j=0}^{d-1} \q_j \ket{j}\bra{j}_M$. Then, we perform
$m$ swaps between the hot virtual reservoir of the passive state and $m$ different pairs of states of $\rho_M$,
followed by $n$ swaps between the cold virtual reservoir and other $n$ different pairs of states of $\rho_M$.
In order to perform the swaps on different pairs of states, we need the machine to have at least $m+n$
levels, and therefore we fix $d = m+n$. Specifically, we apply the following unitary operation to the global
system
\begin{equation} \label{global_evolution}
S_{m,n} = S_{(1,2)}^{(0,m)} \circ S_{(1,2)}^{(m,m+1)} \circ S_{(1,2)}^{(m+1,m+2)} \circ \ldots \circ
S_{(1,2)}^{(m+n-2,m+n-1)} \circ S_{(0,1)}^{(m-1,m+n-1)} \circ S_{(0,1)}^{(m-2,m-1)} \circ
S_{(0,1)}^{(m-3,m-2)} \circ \ldots \circ S_{(0,1)}^{(0,1)},
\end{equation}
where the operator $S_{(a,b)}^{(c,d)}$ is a swap between system and machine, performed
through the permutation $\ket{a}_P \ket{d}_M \leftrightarrow \ket{b}_P \ket{c}_M$. A graphical
representation of this global operation is shown in Fig.~\ref{fig3:catalyst_diag}, where each
swap is depicted by an arrow acting over the states of the machine. Although in the
figure we represent the eigenstates of $\rho_M$ in a ladder, they are all associated with the same
energy, and therefore the order in which we present them is only functional for visualising the cycle
$S_{m,n}$.
\par
In order for the engine to be re-usable, we need the local state of the machine to end up in
its initial state. Therefore, we impose the following constraint on the state of the machine,
\begin{equation} \label{machine_condition}
\rho_M \overset{!}{=} \Tr{P}{S_{m,n} \left( \rho_P \otimes \rho_M \right) S_{m,n}^{\dagger}}.
\end{equation}
Through Eq.~\eqref{machine_condition} we can express the probability distribution of the machine
in terms of the passive state $\rho_P$ (see the Supplemental Material for further details). In our
model we do not explicitly include an additional system (a battery) for storing the energy we extract
from the passive state. Instead, we implicitly assume the existence of this work-storage system, and
we define the work extracted, $\Delta W$, as the difference in average energy between the initial and
final state of the main system (as the machine $M$ has a trivial Hamiltonian, and no interaction
terms are present between system and machine). Thus, we have that
\begin{equation} \label{definition_work}
\Delta W = \tr{H_P \left( \rho_P - \tilde{\rho}_P \right)},
\end{equation} 
where the final state of the system is
\begin{equation} \label{final_state_passive}
\tilde{\rho}_P = \Tr{M}{S_{m,n} \left( \rho_P \otimes \rho_M \right) S_{m,n}^{\dagger}}.
\end{equation}
It is worth noting that the final state of system and machine will in general develop correlations.
These correlations are classical, and without them work would not be
extracted during the cycle. However, they do not compromise the re-usability of the machine if applied
to another uncorrelated quantum system.
\par
\begin{figure}[!ht]
\center
\includegraphics[width=0.45\textwidth]{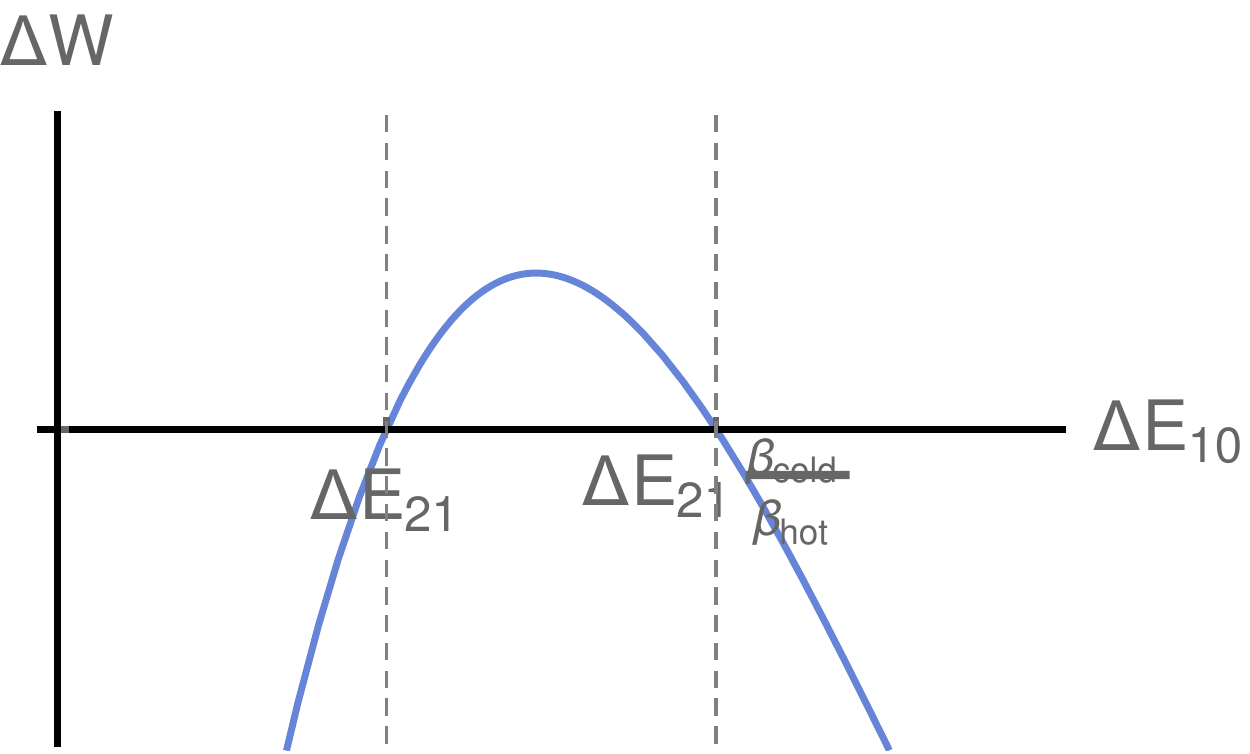}
\includegraphics[width=0.45\textwidth]{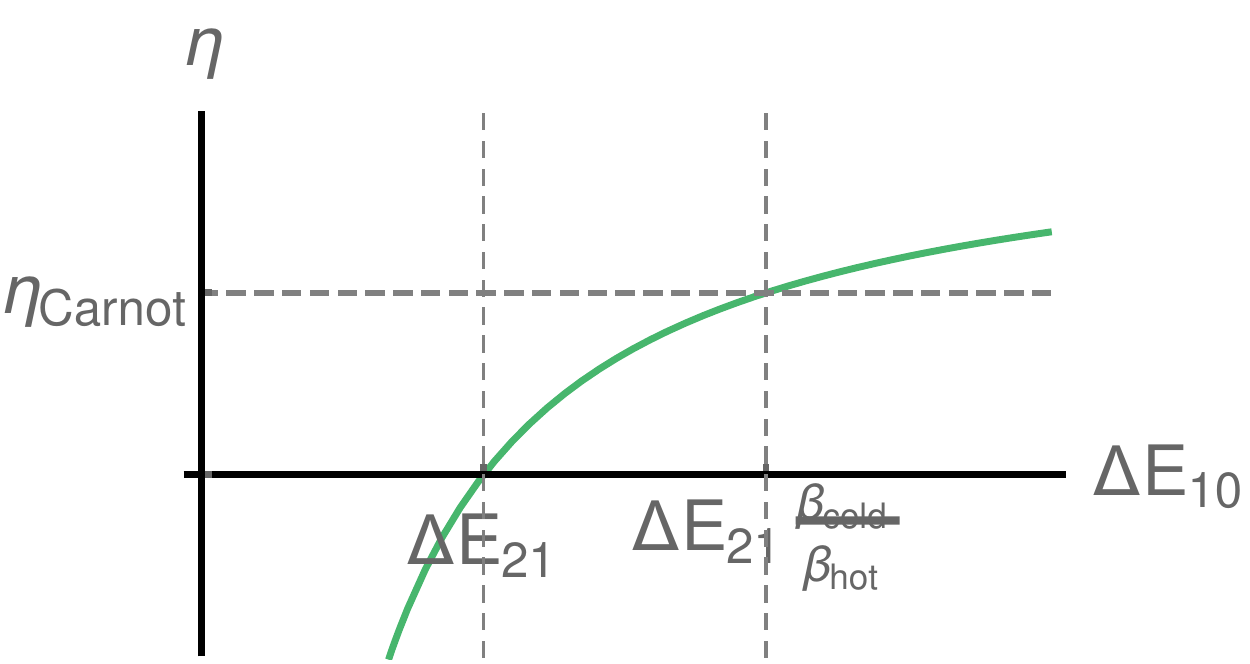}
\caption{We consider the case in which the machine is a qubit system, and we
only perform a single hot and cold swap, that is $m, n = 1$. This cycle is analogous
to the one studied in Ref.~\cite{linden_how_2010}. (Left) The work extracted
in the cycle is positive if the hot energy gap $\Delta E_{10}$ lies inside the range
$\left[ \Delta E_{10} ; \frac{ \beta_{\text{cold}} }{ \beta_{\text{hot}} } \Delta E_{10}
\right]$, as we would expect from conditions~\ref{work_ext_cond_1} and \ref{work_ext_cond_2}.
(Right) The efficiency of the engine, as given by Eq.~\eqref{efficiency}.}
\label{fig4:two_strokes_engine}
\end{figure}

For a given system Hamiltonian $H_P$ and a given cycle $S_{m,n}$ we can investigate the amount
of work we extract from the state $\rho_P$. In the Supplemental Material we provide all the necessary
steps to evaluate $\Delta W$ in terms of the probability distribution of $\rho_P$. We can express this quantity as
\begin{equation} \label{ultimate_work}
\Delta W = \alpha \left( m \, \Delta E_{10} - n \, \Delta E_{21} \right)
\left( e^{\beta_{\text{cold}} n \Delta E_{21}} - e^{\beta_{\text{hot}} m \Delta E_{10}} \right),
\end{equation}
where $\alpha$ is a positive coefficient depending non-trivially on the probability distribution of
$\rho_P$. For the class of passive states we are considering (namely, the one in which
$\beta_{\text{cold}} > \beta_{\text{hot}}$), we find that work can be extracted ($\Delta W > 0$) iff
\begin{enumerate}
\item The Hamiltonian $H_P$ is such that $m \, \Delta E_{10} > n \, \Delta E_{21}$.
\label{work_ext_cond_1}
\item The temperatures of the two virtual reservoirs are such that
$\beta_{\text{cold}} > \frac{m \, \Delta E_{10}}{n \, \Delta E_{21}} \beta_{\text{hot}}$.
\label{work_ext_cond_2}
\end{enumerate}
Thus, for a fixed cycle (defined by the parameters $m$ and $n$), and for a fixed Hamiltonian $H_P$,
we find that work can only be extracted if the virtual temperature $T_{\text{cold}}$ is lower than
$T_{\text{hot}}$ by a multiplicative factor which depends on the energy gaps of the Hamiltonian,
see Fig.~\ref{fig4:two_strokes_engine}, left plot, for an example.
In Sec.~\ref{activation_region} we show that, for a given Hamiltonian $H_P$, work can be extracted
by any passive (but not completely passive) state, and we characterise the cycle which allows for this
extraction.
\par
If we analyse in a more detailed way the cycle, we find that the same
amount of energy is gained during each swap between the machine $M$ and the hot virtual reservoir, that is
\begin{equation} \label{work_hot_swap}
q_{\text{hot}} =  \alpha \, \Delta E_{10}
\left( e^{\beta_{\text{cold}} n \Delta E_{21}} - e^{\beta_{\text{hot}} m \Delta E_{10}} \right),
\end{equation}
where $\alpha$ is the same positive coefficient of Eq.~\eqref{ultimate_work}. Moreover, the same amount
of energy is spent during each swap between the machine $M$ and the cold virtual reservoir,
\begin{equation} \label{work_cold_swap}
q_{\text{cold}} = \alpha \, \Delta E_{21}
\left( e^{\beta_{\text{cold}} n \Delta E_{21}} - e^{\beta_{\text{hot}} m \Delta E_{10}} \right).
\end{equation}
\par
Knowing the amount of energy exchanged during each swap allows us to evaluate the heat exchanged
with the virtual reservoirs. In fact, if we identify the pair of levels $\left( \ket{0}_P , \ket{1}_P \right)$ with the
hot virtual reservoir, then the energy exchanged during a swap with these levels can be considered as heat
coming from the hot virtual reservoir. In this way, the total heat absorbed by the machine is
\begin{equation} \label{heat_hot}
Q_{\text{hot}} = m \, q_{\text{hot}},
\end{equation}
while the total heat provided to the cold virtual reservoir is
\begin{equation} \label{heat_cold}
Q_{\text{cold}} = n \, q_{\text{cold}}.
\end{equation}
From Eqs.~\eqref{heat_hot} and \eqref{heat_cold} we obtain that the work extracted can be
expressed as $\Delta W = Q_{\text{hot}}  - Q_{\text{cold}}$, as in a standard heat engine exchanging
energy between two reservoirs. Once $Q_{\text{hot}}$ and $Q_{\text{cold}}$ are defined, we
can evaluate an efficiency of this cycle, that is
\begin{equation} \label{efficiency}
\eta = \frac{ \Delta W }{ Q_{\text{hot}} } = 1 - \frac{ n \, \Delta E_{21} }{ m \, \Delta E_{10} }.
\end{equation}
The efficiency of the engine (when the machine is finite-dimensional) is sub-Carnot in the virtual temperatures, see
Fig.~\ref{fig4:two_strokes_engine}, right plot. In fact, work
can only be extracted when conditions~\ref{work_ext_cond_1} and \ref{work_ext_cond_2} are satisfied,
and these conditions implied $0 < \eta < 1 - \frac{T_{\text{cold}}}{T_{\text{hot}}}$. When we consider the
case of an infinite-dimensional machine, we find that by a judicious choice of parameters we may obtain Carnot efficiency.
\par
Once the cycle $S_{m,n}$ is ended, the local state of the main system is moved to a less energetic
state. By solving Eq.~\eqref{machine_condition},
we find that the final state of the main system $\tilde{\rho}_P$ has the following probability distribution
\begin{subequations}
\label{final_prob_dist}
\begin{align}
\p'_0 &= \p_0 + m \, \Delta \mathrm{P}, \\
\p'_1 &= \p_1 - \left( m + n \right) \, \Delta \mathrm{P}, \\
\p'_2 &= \p_2 + n \, \Delta \mathrm{P},
\end{align}
\end{subequations}
where the unit of probability $\Delta \mathrm{P}$ depends on the initial state $\rho_P$, and
it is given by
\begin{equation}
\label{prob_unit}
\Delta \mathrm{P} = \alpha
\left( e^{\beta_{\text{cold}} n \Delta E_{21}} - e^{\beta_{\text{hot}} m \Delta E_{10}} \right),
\end{equation}
with $\alpha$ the same positive coefficient of Eq.~\eqref{ultimate_work}. Due to condition
\ref{work_ext_cond_2}, the unit $\Delta \mathrm{P} > 0$, so that the probability of occupation
of $\ket{1}_P$ is reduced in favour of the probabilities $\p_0$ and $\p_1$. Thus, energy is
extracted from the passive state when $m \, \Delta \mathrm{P}$ is moved from $\p_1$ to
$\p_0$ (during the hot swaps), and part of this energy is used to move the probability $n \,
\Delta \mathrm{P}$ from $\p_1$ to $\p_2$ (during the cold swaps).
\section{Work extraction from any passive state}
\label{activation_region}
We now show that, for a given Hamiltonian $H_P$, work can be extracted from any passive
but not completely passive state. In particular, we first show this for qutrit passive states, and
we then generalise to the qudit case. Work extraction is achieved with the cycle presented in
Sec.~\ref{general_cycle}, for specific values of the parameters $m$ and $n$. In what follows,
we represent the passive state with the probabilities of occupation $\left\{ \p_0 , \p_1 , \p_2 \right\}$,
as opposed to the previous case in which the virtual temperatures were used. In this way, we
can consider all possible scenarios, and we are not limited to the case in which a specific pair
of eigenstates has a colder (hotter) virtual temperature than the other pair.
\begin{figure}[!ht]
\center
\includegraphics[width=0.45\textwidth]{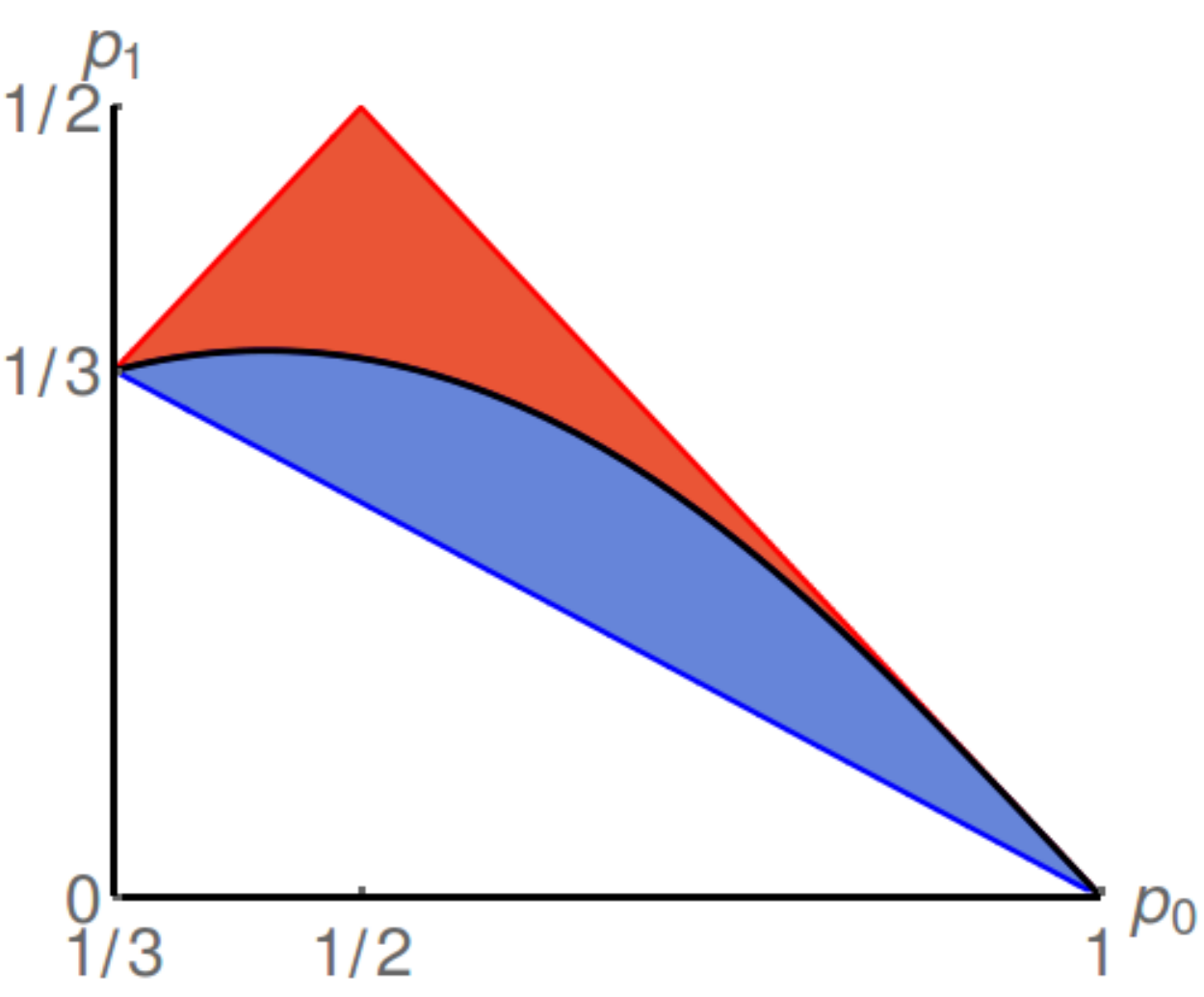}
\includegraphics[width=0.45\textwidth]{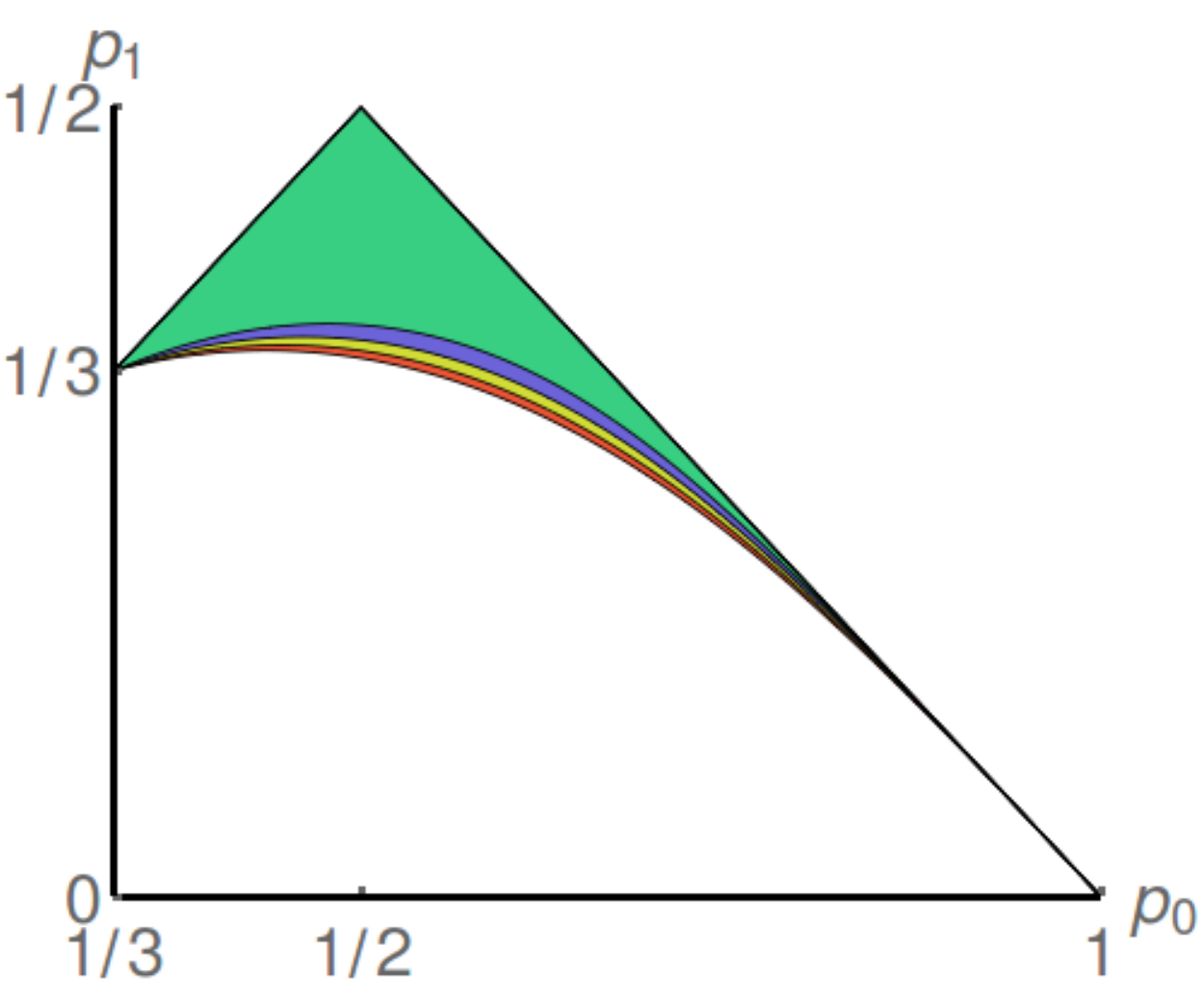}
\caption{(Left) The set of qutrit passive states. The red region represents the subset $R_1$,
while the blue region represents the subset $R_2$. The black line is $R_3$, that is, the set of
completely passive states. (Right) The region $R_1$ (in red) contains the regions $R^{+}_{3,1}$
(green), $R^{+}_{5,2}$ (purple), and $R^{+}_{11,5}$ (yellow), which cover $R_1$ better and better
as $m$ and $n$ grow. In both plots we set $M = 2$ and $N = 1$.}
\label{fig5:passive_region}
\end{figure}
\par
The Hamiltonian of the system $H_P$ is defined in Eq.~\eqref{hamiltonian}, where the energy
gap between ground and first excited state is $\Delta E_{10}$, and the gap between first
and second excited states is $\Delta E_{21}$. We assume that
\begin{equation} \label{hamiltonian_condition}
\exists \, M, N \in \N \ \text{such that} \ M \, \Delta E_{10} - N \, \Delta E_{21} = 0,
\end{equation}
that is, we ask the ratio between the two energy gaps to be rational. Notice that, even if the
ratio is irrational, we can find a suitable $N$ and $M$ such that the condition is approximatively
satisfied. Once the relation between energy gaps is defined, we can divide the set of passive
states into three different subsets, namely
\begin{subequations}
\label{passive_subsets}
\begin{align}
\label{passive_R_1}
R_1 &= \left\{ \rho_P \ \text{passive} \ \Big| \ \left( \frac{\p_1}{\p_2} \right)^N > \left( \frac{\p_0}{\p_1} \right)^M  \right\}, \\
\label{passive_R_2}
R_2 &= \left\{ \rho_P \ \text{passive} \ \Big| \ \left( \frac{\p_1}{\p_2} \right)^N < \left( \frac{\p_0}{\p_1} \right)^M  \right\}, \\
\label{passive_R_3}
R_3 &= \left\{ \rho_P \ \text{passive} \ \Big| \ \left( \frac{\p_1}{\p_2} \right)^N = \left( \frac{\p_0}{\p_1} \right)^M  \right\}.
\end{align}
\end{subequations}
The union of these three subsets gives the set of all passive states. In particular, one can verify that
the subset $R_3$ contains all the completely passive states, that is, the thermal states of $H_P$ at
any temperature $\beta^{-1} \geq 0$. Moreover, $R_1$ corresponds to the set of passive states with
$\beta_{\text{hot}}$ associated with the pair of eigenstates $\ket{0}_P$ and $\ket{1}_P$, and
$\beta_{\text{cold}}$ associated with the pair $\ket{1}_P$ and $\ket{2}_P$. The set $R_2$, instead,
contains the passive states with opposite hot and cold virtual temperatures.
Since we are considering qutrit systems, we can represent the set of passive states
in a two-dimensional diagram, using their probability distribution. Each point in this diagram represents a
passive state. In Fig.~\ref{fig5:passive_region}, left plot, we show the three subsets of Eqs.~\eqref{passive_subsets}.
\par
In the previous section, we have seen that a cycle defined by the parameters $m$ and $n$ can
activate a passive state $\rho_P$ with Hamiltonian $H_P$ if conditions \ref{work_ext_cond_1}
and \ref{work_ext_cond_2} are satisfied. These conditions apply to the case in which the passive
state is described by Eqs.~\ref{hot_cold_temperatures}, with $\beta_{\text{hot}} < \beta_{\text{cold}}$.
In the present, more general scenario we find that work is extracted by the cycle if and only if the
passive state belongs to the following subset
\begin{align} \label{activable_region_with_cycle}
R^{+}_{m,n} = \bigg\{ \rho_P \ \text{passive} \ \bigg| 
\quad &\left( \frac{\p_1}{\p_2} \right)^n > \left( \frac{\p_0}{\p_1} \right)^m
\text{when} \
m \, \Delta E_{10} - n \, \Delta E_{21} > 0 \nonumber \\
\vee
\quad &\left( \frac{\p_1}{\p_2} \right)^n < \left( \frac{\p_0}{\p_1} \right)^m
\text{when} \
m \, \Delta E_{10} - n \, \Delta E_{21} < 0
\bigg\},
\end{align}
where these conditions can be obtained by analysing the general expression of the extracted work,
see Supplemental Material. 
We now show that, by tailoring the value of the parameters $m$ and $n$, we can make $R^{+}_{m,n}$
to (asymptotically) cover either the region $R_1$ or $R_2$. Here, we focus on $R_1$ solely, since $R_2$
follows from similar arguments. As a first step, we ask $m \, \Delta E_{10} - n \, \Delta E_{21} > 0$, which
implies $m > \frac{M}{N} n$, due to Eq.~\eqref{hamiltonian_condition}. Then, in order to satisfy this condition,
we set $m = \frac{M}{N} n + 1$, where we ask $n$ to be large enough for $m$ to be an integer. The set
of passive states activated by the cycle is such that
\begin{equation} \label{region_cond_1}
\left( \frac{\p_1}{\p_2} \right)^n > \left( \frac{\p_0}{\p_1} \right)^m
\Rightarrow
\left( \frac{\p_1}{\p_2} \right)^n > \left( \frac{\p_0}{\p_1} \right)^{\frac{M}{N} n + 1}
\Rightarrow
\left( \frac{\p_1}{\p_2} \right)^N > \left( \frac{\p_0}{\p_1} \right)^{M + \frac{N}{n}}.
\end{equation}
We notice that, since $\rho_P$ is passive, $\p_0 \geq \p_1$, which implies that
\begin{equation} \label{region_cond_2}
\left( \frac{\p_0}{\p_1} \right)^{M + \frac{N}{n}} \geq \left( \frac{\p_0}{\p_1} \right)^M .
\end{equation}
Thus, Eq.~\eqref{region_cond_1} and \eqref{region_cond_2} together assure that
$R^{+}_{\frac{M}{N} n + 1,n} \subset R_1$. Moreover, if $n \rightarrow \infty$, we
have that $M + \frac{N}{n} \rightarrow M$,  which implies that $R^{+}_{\frac{M}{N} n + 1,n}
\rightarrow R_1$. Thus, we have that, for a given Hamiltonian $H_P$, and a given passive
state $\rho_P \in R_1$, there exist a cycle $S_{m,n}$ such that $\rho_P \in R^{+}_{m,n}$.
However, the closer (in trace norm) the state $\rho_P$ is to the set of completely passive
states ($R_3$), the larger the parameters $m$ and $n$ have to be, that is, the larger the
machine has to be (see Fig.~\ref{fig5:passive_region}, right plot).
\subsection{Work extraction from a generic qudit passive state}
Work extraction from a generic qudit passive state $\rho_P^{(d)}$ (for any Hamiltonian $H_P^{(d)}$)
can be achieved with the cycle introduced in Sec.~\ref{general_cycle}, even if this work extraction is
not optimal (as it might be when we deal with qutrit state, as we see in Sec.~\ref{asympt_machine}).
Indeed, even if the system has $d$ levels, we only need to focus our analysis on three of them, and
perform the cycle on these levels only. Thus, given the state  $\rho_P^{(d)} = \sum_{i=0}^{d-1} \p_i
\ket{i}\bra{i}_P$ and the Hamiltonian $H_P^{(d)} = \sum_{i=0}^{d-1} E_i \ket{i}\bra{i}_P$, we can
consider the subspace $A_k = \text{span} \left\{ \ket{k}_P , \ket{k+1}_P , \ket{k+2}_P \right\}$, for
a given $k \in [0,d-3]$. Thus, we can divide the qudit state and the Hamiltonian in two contributions,
one with support over $A_k$, the other with support over its complement,
\begin{subequations}
\begin{align}
\rho_P^{(d)} &= \left( \sum_{i \in A_k} \p_i \right) \rho_P^{(A)} + \left( 1 - \sum_{i \in A_k} \p_i \right) \rho_P^{(A^c)}, \\
H_P^{(d)} &= H_P^{(A)} + H_P^{(A^c)},
\end{align}
\end{subequations}
where the normalised quantum states are 
\begin{subequations}
\begin{align}
\rho_P^{(A)} &= \sum_{i \in A_k} \frac{\p_i}{\sum_{j \in A_k} \p_j} \ket{i}\bra{i}_P, \\
\rho_P^{(A^c)} &= \sum_{i \notin A_k} \frac{\p_i}{1 - \sum_{j \in A_k} \p_j} \ket{i}\bra{i}_P,
\end{align}
\end{subequations}
while the Hamiltonian contributions are, respectively, $H_P^{(A)} = \sum_{i \in A_k} E_i
\ket{i}\bra{i}_P$ and $H_P^{(A^c)} = \sum_{i \notin A_k} E_i \ket{i}\bra{i}_P$. In the following,
we define $\lambda = \sum_{i \in A_k} \p_i$, so that $\rho_P^{(d)} = \lambda \, \rho_P^{(A)} +
\left( 1 - \lambda \right) \rho_P^{(A^c)}$.
\par
We can now introduce an ancillary system (the machine $M$) of dimension $m+n$, described
by the state $\rho_M$, together with the global unitary operator $U$,
\begin{equation}
U = P_{A^c} \otimes \Id_M + P_{A} \otimes \Id_M \circ S_{m,n} \circ P_{A} \otimes \Id_M,
\end{equation}
where the operator $S_{m,n}$, described in Eq.~\eqref{global_evolution}, has support on
$A_k \otimes \hil_M$, and therefore commute with $P_{A} \otimes \Id_M$.
If we consider the evolution of the system under this operator, we obtain
\begin{equation}
\tilde{\rho}_P^{(d)} = \Tr{M}{U \left( \rho_P^{(d)} \otimes \rho_M \right) U^{\dagger}}
= \lambda \, \Tr{M}{S_{m,n} \left( \rho_P^{(A)} \otimes \rho_M \right) S_{m,n}^{\dagger}}
+ \left( 1 - \lambda \right) \rho_P^{(A^c)} = \lambda \, \tilde{\rho}_P^{(A)} + \left( 1 - \lambda
\right) \rho_P^{(A^c)},
\end{equation}
and we can easily verify, due to the properties of $S_{m,n}$, that the local state of the machine
is left unchanged. The amount of work extracted during this cycle is
\begin{equation}
\Delta W = \Tr{P}{ H_P^{(d)} \left( \rho_P^{(d)} - \tilde{\rho}_P^{(d)} \right) } =
\lambda \, \Tr{P}{ H_P^{(A)} \left( \rho_P^{(A)} - \tilde{\rho}_P^{(A)} \right) },
\end{equation}
and the problem reduces to the one analysed at the beginning of this section (that is, to the
extraction of work from a qutrit system described by the passive state $\rho_P^{(A)}$, with
Hamiltonian $H_P^{(A)}$), with the only difference of a multiplicative factor $\lambda \in
(0,1)$ in $\Delta W$.
\subsection{Work extraction and \texorpdfstring{$k$}{k}-activable states}
The set of passive states can be divided into a hierarchy of classes, which divides the states
according to the number of copies needed to activate them. Here, we say that a state is \emph{active}
if it is not passive, and therefore if we can extract work from it with unitary operations. Any
passive but not completely passive state can be activated if we tensor together enough
copies of it. In particular, when $k$ copies of a passive state are active, we call the state
\emph{$k$-activable}. We now show that, if work is extracted from a qutrit passive state
$\rho_P$, with Hamiltonian $H_P$, through the cycle $S_{m,n}$, then the state realised
by $m+n$ copies of $\rho_P$ is active. It worth noting that, while our cycle only requires an
additional system of dimension $m+n$ to extract work from $\rho_P$, in order to activate the
same state we would need $m+n-1$ copies of it, that is, an ancilla whose size is exponential
in $n+m$.
\par
In the following, we consider a qutrit system, although the same argument applies to qudit
systems, for the reasons presented in the previous section. If the passive state $\rho_P$ is
activated by the cycle $S_{m,n}$, then one of the two conditions in
Eq.~\eqref{activable_region_with_cycle} has to be satisfied. Let us assume that the conditions
satisfied by state and Hamiltonian are
\begin{subequations}
\label{active_passive_condition}
\begin{align}
m \, \Delta E_{10} &> n \, \Delta E_{21},  \label{active_passive_condition_1} \\
\left( \frac{\p_1}{\p_2} \right)^n &> \left( \frac{\p_0}{\p_1} \right)^m, \label{active_passive_condition_2}
\end{align}
\end{subequations}
where the other case follows straightforwardly.
\par
Consider now a system composed by $n+m$ copies of the qutrit system under examination,
with Hamiltonian $H_{\text{tot}} = \sum_{i=1}^{m+n} H_P^{(i)}$, where the term $H_P^{(i)}$
acts over the $i$-th copy. The state of this global system is $\rho_P^{\otimes m+n}$. Then,
let us focus our attention on two eigenstates of $H_{\text{tot}}$, namely, $\ket{1}^{\otimes m+n}$
and $\ket{0}^{\otimes m} \otimes \ket{2}^{\otimes n}$. The first eigenstate has an energy of $\left(
m + n \right) E_1$, and its occupation probability is $\p_1^{m+n}$. The second eigenstate, instead,
has energy $m \, E_0 + n \, E_2$, and its occupation probability is $\p_0^{m} \p_2^{n}$. It is easy
to verify that, if the constraints of Eqs.~\eqref{active_passive_condition} hold, then the inequalities
$\left( m + n \right) E_1 > m \, E_0 + n \, E_2$ and $\p_1^{m+n} > \p_0^{m} \p_2^{n}$
are satisfied, implying that the state $\rho_P^{\otimes m+n}$ is active.
Thus, we have shown that if a passive state $\rho_P$ can be activated with the cycle $S_{m,n}$,
then the state $\rho_P^{\otimes m+n}$ is active. However, this result does not tell us whether
it is possible to activate the state $\rho_P$ by tensoring it with less copies. In the same way,
we do not know whether the fact that the state $\rho_P^{\otimes m+n}$ is active implies that
we can extract work from $\rho_P$ with the cycle $S_{m,n}$.
\section{General Instability of Passive States}
\label{asympt_machine}
We can now establish our central claim: that any athermal passive state is energetically unstable
under a reversible process that does not generate entropy. We analyse the evolution of a passive state
which sequentially interacts with an infinite-dimensional machine $M$, and find that the system moves
through a continuous trajectory of passive states towards the set of minimum-energy states, that is, the
set of the states~\cite{jaynes_information_1957}.
\par
We consider a cycle composed of infinitely many hot swaps, $m \rightarrow \infty$, and infinite many
cold swaps, $n \rightarrow \infty$, with the assumption that $n = \alpha \, m$, where $\alpha$ is a
parameter taking values in a specific range we will describe shortly. Let us now consider the situation
in which the main system is a qutrit with Hamiltonian $H_P$ given in Eq.~\eqref{hamiltonian},
described by the passive state $\rho_P$ whose probability distribution satisfies the equalities of 
Eqs.~\eqref{hot_cold_temperatures}. Then, $\rho_P$ belongs to the subset $R_1$ defined in
Eq.~\eqref{passive_R_1}, and the cycle $S_{m,n}$ has to satisfy conditions \ref{work_ext_cond_1}
and \ref{work_ext_cond_2} in order to extract work from it. These conditions are reflected
in the allowed range of the parameter $\alpha$, that is
\begin{equation}
\frac{\beta_{\text{hot}} \Delta E_{10}}{\beta_{\text{cold}} \Delta E_{21}}
<
\alpha
<
\frac{\Delta E_{10}}{\Delta E_{21}}.
\end{equation} 
\par
If we set $\alpha$ equal to a value inside the range specified by the previous equation, and we
send $m \rightarrow \infty$, we find that (see Supplemental Material for details) the state of the machine
as obtained from Eq.~\eqref{machine_condition} is given by a mixture of two ``thermal" states,
one with effective temperature $\beta_{\text{hot}}^{-1}$, the other with effective temperature
$\beta_{\text{cold}}^{-1}$ (note we still have $H_M=0$ for the machine). These distributions have support
in two different subspaces, and their weight depends non-trivially on the energy gaps of $H_P$ and on the virtual
temperatures of $\rho_P$. In fact, we can loosely interpret the state of the machine in terms of a thermal mixture
\begin{equation}
\rho_M = \lambda \, \tau_{\beta_{\text{hot}}} + \left( 1 - \lambda \right) \tau_{\beta_{\text{cold}}},
\end{equation}
where
\begin{equation}
\tau_{\beta_{\text{hot}}} = \frac{ e^{-\beta_{\text{hot}} H_{\text{hot}} } }{ Z_{\text{hot}} }
\ , \
\text{with}
\
H_{\text{hot}} = \sum_{j=0}^{m-1} j \, \Delta E_{10} \ket{j}\bra{j}_M
\
\text{and}
\
Z_{\text{hot}} = \tr{ e^{-\beta_{\text{hot}} H_{\text{hot}} } },
\end{equation}
and
\begin{equation}
\tau_{\beta_{\text{cold}}} = \frac{ e^{-\beta_{\text{cold}} H_{\text{cold}} } }{ Z_{\text{cold}} }
\ , \
\text{with}
\
H_{\text{cold}} = \sum_{j=0}^{n-3} j \, \Delta E_{21} \ket{j+m}\bra{j+m}_M
\
\text{and}
\
Z_{\text{cold}} = \tr{ e^{-\beta_{\text{cold}} H_{\text{cold}} } }.
\end{equation}
Notice that in order to define these ``thermal states" we have introduced two fictitious Hamiltonians, namely,
$H_{\text{hot}}$ and $H_{\text{cold}}$. These operators are necessary if we want to consider the
distribution of the machine as the mixture of two thermal distributions, but they do not enter in any
way in the derivation of the extractable work. Indeed, as we have specified at the beginning of
Sec.~\ref{general_cycle}, the machine $M$ can have any Hamiltonian (it does not modify the
amount of work we extract during the cycle), and we choose to use a trivial one $H_M=0$, so that the machine
acts as a memory. The weight $\lambda$ in the mixture is given by
\begin{equation}
\lambda = \frac{ 1 - e^{ -\beta_{\text{cold}} \Delta E_{21} } }
{ 1 - e^{ -\beta_{\text{hot}} \Delta E_{10} - \beta_{\text{cold}} \Delta E_{21} } }.
\end{equation}
Thus, during the cycle, the passive state $\rho_P$ first interacts with the ``hot reservoir", by
performing a sequence of swaps between the pair of states $\ket{0}_P$ and $\ket{1}_P$ and the
levels of $\tau_{\beta_{\text{hot}}}$. Then, the state interacts with the ``cold reservoir", performing
a sequence of swaps between the pair $\ket{1}_P$ and $\ket{2}_P$ and the levels of
$\tau_{\beta_{\text{cold}}}$.
\par
\begin{figure}[ht!]
\center
\includegraphics[width=0.45\textwidth]{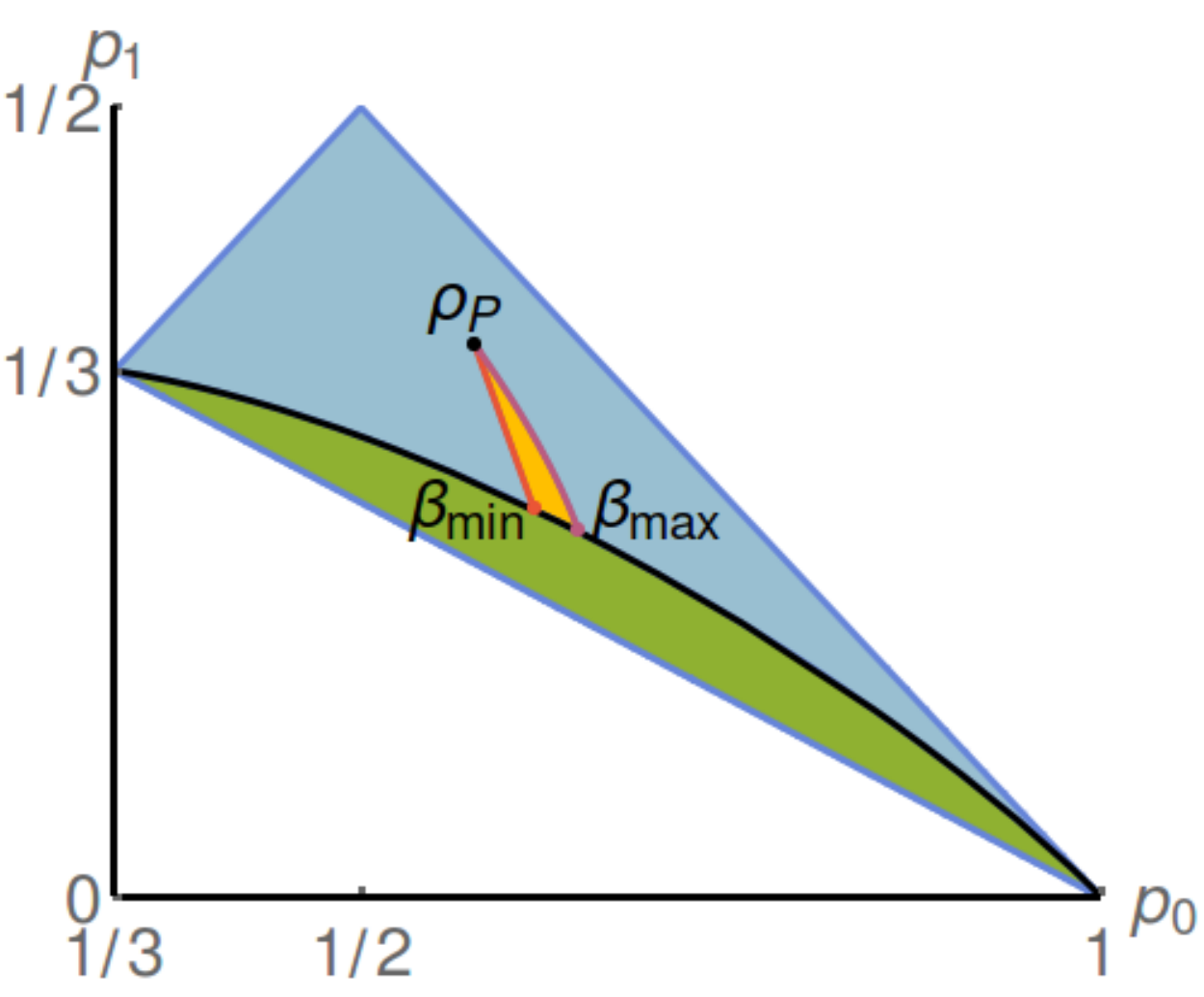}
\includegraphics[width=0.45\textwidth]{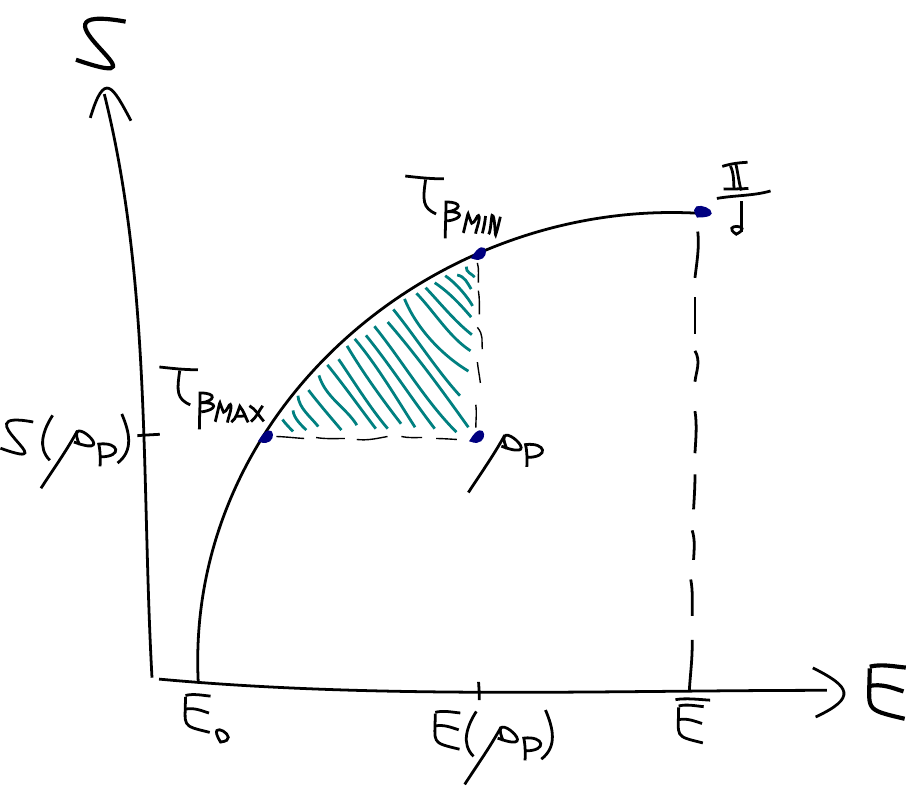}
\caption{(Left) The state space of a qutrit system, where the subset of passive states is highlighted. The blue region is
$R_1$, while the green region is $R_2$. The black line is the set of thermal states. The initial state $\rho_P$
is represented by the black point in the diagram. If $\alpha$ takes the value $\frac{\Delta E_{10}}
{\Delta E_{21}}$, then the system evolves along the red trajectory, and the final state is the thermal state
at temperature $\beta_{\text{min}}$. On the other hand, if $\alpha = \frac{\beta_{\text{hot}}(t) \Delta E_{10}}
{\beta_{\text{cold}}(t) \Delta E_{21}}$, the system evolves along the purple line, and the final state is the
thermal state at temperature $\beta_{\text{max}}$. The yellow region represents the subset of achievable states 
when the initial state is $\rho_P$. (Right) A partial representation of the state space of a $d$-level quantum system
in the energy-entropy diagram. In this diagram, quantum states are grouped into equivalence classes defined by their
average energy $E$ and their entropy $S$. Each point between the $x$-axis (the set of pure states) and the black
curve (the set of thermal states) represents one of these equivalence classes. The diagram depends on the Hamiltonian $H_P$
of the system. Here, we only represent the states with average energy lower than  $\bar{E} = \tr{ H_P
\rho_{\text{mm}}}$, where $\rho_{\text{mm}} = \frac{\Id}{\mathrm{d}}$ is the maximally-mixed state,
since all passive states are contained in this set. For a given initial state $\rho_P$, the green
region contains all the passive states which can be achieved with the process.
}
\label{fig:asympt_evolution}
\end{figure}

In this scenario, we find that the probability distribution of the passive state $\rho_P$ is infinitesimally
modified, and consequently the work extracted is infinitesimally small. In particular, we find that the
unit of probability, defined in Eq.~\eqref{prob_unit}, tends to $0$ with an exponential scaling,
$\Delta \mathrm{P} \propto e^{-\beta_{\text{hot}} m \Delta E_{10} }$ for $m \rightarrow \infty$.
Let us consider the probability distribution of the final state of the system $\tilde{\rho}_P$.
Since the distribution only changes infinitesimally during the cycle, we can recast Eqs.~\eqref{final_prob_dist}
as a set of differential equations (see the Supplemental Material for further details).
Thus, we can imagine the situation in which infinite many machines are present,
so that we can keep infinitesimally changing the state of the main system.
In this case, the evolution of the state $\rho_P$ is governed by the following equation
\begin{equation}
\label{diff_eq}
\frac{ \rmd \p_1 }{ \rmd t } = - \Big( 1 + \alpha \big( p_0(t) , p_1(t) \big) \Big) \frac{ \rmd \p_0 }{ \rmd t },
\end{equation}
where the parameter $t$ provides a continuum label for the sequence of cycles we perform on the passive state.
We can then solve this equation for extremal cases for the function $\alpha(p_0,p_1)$. When the parameter function
$\alpha(p_0,(t),p_1(t))$ is equal to one of its limiting values, Eq.~\eqref{diff_eq} assumes a clear meaning. In fact,
\begin{itemize}
\item when $\alpha(p_0(t),p_1(t)) = \frac{\Delta E_{10}}{\Delta E_{21}}$, then the differential equation can
be recast as a condition over the average energy of the system, that is,
\begin{equation}
\tr{ H_P \, \rho_P} = \tr{ H_P \, \tilde{\rho}_P}.
\end{equation}
Then, for $\alpha$ taking this value, the passive state evolves along a trajectory that conserves the energy
of the system.
\item when $\alpha(p_0(t),p_1(t)) = \frac{\beta_{\text{hot}}(t) \Delta E_{10}}{\beta_{\text{cold}}(t) \Delta E_{21}}$,
instead, the differential equation can be recast as a condition over the entropy of the system, that is,
\begin{equation}
S \left( \rho_P \right) = S \left( \tilde{\rho}_P \right),
\end{equation}
where $S(\rho) = - \tr{\rho \log \rho}$ is the Von Neumann entropy. Then, for this $\alpha$, the
passive state evolves along a trajectory that conserves the entropy of the system.
\end{itemize}
For $\alpha$ taking values inside the allowed range, we have that any trajectory between the two presented
above is possible, and the set of achievable states is shown in Fig.~\ref{fig:asympt_evolution}, left plot.
It is possible to show that the evolution of the system moves the passive state toward the set of thermal
states, which are the stationary states of this dynamic. In Fig.~\ref{fig:asympt_evolution}, right plot, we
show the same set of achievable states, represented this time in the energy-entropy
diagram~\cite{sparaciari_resource_2016}. It is clear that, through this evolution, we can obtain any passive
state with a smaller average energy and a bigger entropy than $\rho_P$. In Supplemental Material, we show that these
states are also the only ones that we can reach with our engines (and with a broader class of maps,
called \emph{activation maps}).
\par
It is interesting to consider the limiting values of work extraction that can be achieved following the scheme
suggested in this section. In particular, when the system evolves along the energy-preserving trajectory, the
final state we obtain is the thermal state of $H_P$ at temperature $\beta_{\text{min}}^{-1}$, that is,
$\tau_{\beta_{\text{min}}}$, where the temperature is such that $\tr{ H_P \, \rho_P} = \tr{ H_P \,
\tau_{\beta_{\text{min}}}}$. In this case, it is easy to see that the engine is not extracting any work, and its only
effect consists in raising the entropy of the system. If we consider the efficiency of this cycle, Eq.~\eqref{efficiency},
we see that $\eta = 0$, as expected. The opposite limit is obtained when the system evolves along the entropy-preserving
trajectory. In this case, the final state is $\tau_{\beta_{\text{max}}}$, that is, the thermal state of $H_P$ at
temperature $\beta_{\text{max}}^{-1}$, such that $S \left( \rho_P \right) = S \left( \tau_{\beta_{\text{max}}} \right)$.
In this case, the work extracted by the cycle is
\begin{align}
  \Delta W = \tr{ H_P \left( \rho_P - \tau_{\beta_{\text{max}}} \right) }\,\, ,
  \label{eq:optimalwork}
  \end{align}
that is the maximum amount one can extract~\cite{alicki_entanglement_2013, sparaciari_resource_2016}.
Significantly, in this case the efficiency is equal to the Carnot one, $\eta_{\text{Carnot}} = 1 - \frac{\beta_{\text{hot}}}
{\beta_{\text{cold}}}$.
\section{Conclusion}
\label{conclusion}
In the paper we have presented an engine that is able to extract work from any single copy of an athermal
passive state. The
engine utilises an ancillary system for the work extraction, and the local state of this system is recovered at
the end of the cycle. In this way, the cycle can be run multiple times, and each time it acts on a new copy of the
passive state. We show that, for any given Hamiltonian, work can be extracted from any passive, but not
completely passive state. Moreover, we show that, in order to extract work from passive states close to the
set of completely passive states, we need an ancillary system of large dimension. With an infinite dimension
machine we can also evolve a passive state smoothly toward the set of thermal states. In particular, optimal
work extraction can be obtained in this case, and it is achieved by mapping the initial state into the thermal state
with the same entropy.
\par
The present work provides some evidence that a resource theory for thermodynamics with an imperfect thermal
reservoir presents non-trivial challenges. Such a resource theory could be realised by providing passive states for
free. However, an obvious restriction we should make in this resource theory consists in the fact that we could
not provide more than $k-1$ copies of a $k$-activatable passive state, otherwise work might be extracted with
unitary operations from this free state. Moreover, our results show that, even in the case in which a single passive
state is provided, an ancillary system exists such that work can be extracted from the individual passive state.
Then, in order to build a sensible resource theory, passive states should be always provided at a work cost, equal to
the optimal amount of energy extractable from them when a machine is present.
\par
It might also be interesting to analyse which passive states allow for the extraction of the highest
amount of work in our cycle. In particular, this problem has been studied in the case of passive states and
unitary evolution~\cite{perarnau-llobet_most_2015}, and a comparison between that class of states and the
class of states which allow for maximal work extraction during our cycle might be interesting.

\noindent
    {\bf Acknowledgements:} We are
    grateful to Ben Schumacher and Michael Westmoreland for inspiring discussions. 
    JO thanks the Royal Society and an EPSRC Established Career
    Fellowship for their support. CS is supported by the EPSRC [grant number EP/L015242/1].
    DJ is supported by the
    Royal Society. We thank the COST Network MP1209 in Quantum Thermodynamics.

\bibliographystyle{ieeetr}
\bibliography{./biblio_passive}
\clearpage
\onecolumngrid
\section*{Supplemental Material}
\appendix
\section{General cycle for work extraction from passive states}
\label{sec_general_cycle}
We present in full details the general cycle needed to extract work from a qutrit system described by a
passive state. Work extraction is achieved through the interaction between a qudit ancilla (the thermal
machine) and the main qutrit system. This qutrit system has Hamiltonian
\begin{equation}
H_P = E_0 \ket{0}\bra{0}_P + E_1 \ket{1}\bra{1}_P + E_2 \ket{2}\bra{2}_P,
\end{equation}
and we define the energy gap as $\Delta E_{10} = E_1 - E_0 > 0$ and $\Delta E_{21} = E_2 - E_1 > 0$.
The state of the system is passive, meaning that no energy can be extracted with unitary operations,
and we can write it as a classical state
\begin{equation}
\rho_P = \p_0 \ket{0}\bra{0}_P + \p_1 \ket{1}\bra{1}_P + \p_2 \ket{2}\bra{2}_P,
\end{equation}
where $\p_0 \geq \p_1 \geq \p_2$ (which is a direct consequence of the no-energy-extraction condition).
\par
The machine we introduce is a $d$-level system with a trivial Hamiltonian, described by the state
\begin{equation}
\rho_M = \sum_{j=0}^{d-1} \q_j \ket{j}\bra{j}_M.
\end{equation}
We operate over system and machine with a unitary operation composed by multiple
swaps. In particular, we first perform $m-1$ swaps between the
pair of states $\left( \ket{0}_P , \ket{1}_P \right)$ and the pairs $\left\{\left( \ket{j}_M , \ket{j+1}_M
\right) \right\}_{j=0}^{m-2}$, followed by a swap between the same pair of states of the system and
the pair $\left( \ket{m-1}_M , \ket{m+n-1}_M \right)$ of the machine. Then, we perform $n-1$
swaps between the pair $\left( \ket{1}_P , \ket{2}_P \right)$ and the pairs $\left\{ \left( \ket{j}_M ,
\ket{j+1}_M \right) \right\}_{j=m}^{m+n-2}$, followed by a swap between the same system's states
and the pair $\left( \ket{0}_M , \ket{m}_M \right)$. In order to perform this cycle, the dimension
of the catalyst has to be at least equal to $m +n$, and indeed in the following we fix $d = m + n$.
The unitary we want to apply is
\begin{equation}
S_{m,n} = S_{(1,2)}^{(0,m)} \circ S_{(1,2)}^{(m,m+1)} \circ S_{(1,2)}^{(m+1,m+2)} \circ \ldots \circ
S_{(1,2)}^{(m+n-2,m+n-1)} \circ S_{(0,1)}^{(m-1,m+n-1)} \circ S_{(0,1)}^{(m-2,m-1)} \circ
S_{(0,1)}^{(m-3,m-2)} \circ \ldots \circ S_{(0,1)}^{(0,1)},
\end{equation}
where the operation $S_{(a,b)}^{(c,d)}$ is a swap between system and machine, performing
the permutation $\ket{a}_P \ket{d}_M \leftrightarrow \ket{b}_P \ket{c}_M$.
\par
For the given unitary evolution we can easily evaluate the final state of the global system.
This final state presents classical correlations between system and machine, but in
the following we only consider the marginal states for system and machine, which are the
sole information we need. In fact, the energy of the global system solely depends on the
Hamiltonian $H_P$ of the system (and therefore only on the local state of the system), as
the machine has a trivial Hamiltonian, and we do not have an interaction term
$H_{\text{int}}$. Moreover, in order for the machine to be re-usable on a new system,
we only need its local initial and final states to be equal, and the correlations with the old systems
do not affect the engine. The final state of the system is
\begin{align} \label{final_state_system}
\tilde{\rho}_P &= \Tr{M}{S_{m,n} \left( \rho_P \otimes \rho_M \right) S_{m,n}^{\dagger}} \nonumber \\
&= \left( \p_0 + \sum_{j=1}^{m-1} \left( \p_1 \q_{j-1} - \p_0 \q_j \right) + \left( \p_1 \q_{m-1} - \p_0 \q_{m+n-1} \right) \right)
\ket{0}\bra{0}_P \nonumber \\
&+
\left( \p_1 - \sum_{j=1}^{m-1} \left( \p_1 \q_{j-1} - \p_0 \q_j \right) - \left( \p_1 \q_{m-1} - \p_0 \q_{m+n-1} \right)
- \sum_{j=m+1}^{m+n-1} \left( \p_1 \q_j - \p_2 \q_{j-1} \right) - \left( \p_1 \q_m - \p_2 \q_0 \right) \right)
\ket{1}\bra{1}_P \nonumber \\
&+
\left( \p_2 + \sum_{j=m+1}^{m+n-1} \left( \p_1 \q_j - \p_2 \q_{j-1} \right) + \left( \p_1 \q_m - \p_2 \q_0 \right) \right)
\ket{2}\bra{2}_P,
\end{align}
while the final state of the machine is
\begin{align}
\tilde{\rho}_M &= \Tr{P}{S_{m,n} \left( \rho_P \otimes \rho_M \right) S_{m,n}^{\dagger}} \nonumber \\
&= \left( \p_0 \q_0 + \p_0 \q_1 + \p_1 \q_m \right) \ket{0}\bra{0}_M
+ \sum_{j=1}^{m-2} \left( \p_1 \q_{j-1} + \p_0 \q_{j+1} + \p_2 \q_j \right) \ket{j}\bra{j}_M \nonumber \\
&+ \left( \p_1 \q_{m-2} + \p_0 \q_{m+n-1} + \p_2 \q_{m-1} \right) \ket{m-1}\bra{m-1}_M
+ \left( \p_0 \q_m + \p_2 \q_0 + \p_1 \q_{m+1} \right) \ket{m}\bra{m}_M \nonumber \\
&+ \sum_{j=m+1}^{m+n-2} \left( \p_0 \q_j + \p_2 \q_{j-1} + \p_1 \q_{j+1} \right) \ket{j}\bra{j}_M
+ \left( \p_1 \q_{m-1} + \p_2 \q_{m+n-2} + \p_2 \q_{m+n-1} \right) \ket{m+n-1}\bra{m+n-1}_M.
\end{align}
\par
As we stated above, in order for the machine to be re-usable we need its final local state
$\tilde{\rho}_M$ to be equal to the initial one $\rho_M$. Correlations with the system do not invalidate
the re-usability, as we always discard the system after the cycle, and we take a new copy
to repeat the process. In this way, we can extract work from a reservoir of passive states by
acting on them individually. The constraint of an equal initial and final state of the machine provides the
following set of equalities, 
\begin{subequations}
\label{prob_dist_machine}
\begin{align}
\q_0 &= \p_0 \q_0 + \p_0 \q_1 + \p_1 \q_m \label{prob_dist_1} \\
\q_j &=  \p_1 \q_{j-1} + \p_0 \q_{j+1} + \p_2 \q_j \quad ; \quad j = 1 , \ldots , m - 2 \label{prob_dist_2} \\
\q_{m-1} &= \p_1 \q_{m-2} + \p_0 \q_{m+n-1} + \p_2 \q_{m-1} \label{prob_dist_3} \\
\q_m &= \p_0 \q_m + \p_2 \q_0 + \p_1 \q_{m+1} \label{prob_dist_4} \\
\q_j &= \p_0 \q_j + \p_2 \q_{j-1} + \p_1 \q_{j+1} \quad ; \quad j = m+1 , \ldots ,m+n-2 \label{prob_dist_5} \\
\q_{m+n-1} &= \p_1 \q_{m-1} + \p_2 \q_{m+n-2} + \p_2 \q_{m+n-1}, \label{prob_dist_6}
\end{align}
\end{subequations}
which, if solved, allow for the probability distribution of the state of the machine to be expressed in terms
of the passive state $\rho_P$.
\subsection{Work extracted and activable passive states}
In our framework, we do not explicitly account for a battery, that is, an additional system
with a specific Hamiltonian, able to account for any energy exchange between system and
machine. Instead, we implicitly assume the battery to be present, so that any change in the
average energy of the system is thought as some energy flowing from (or to) the battery.
In particular, if the average energy of the system decreases, then the battery is storing this
energy, while when the average energy of the system increases, the battery is providing it.
All the energy coming from (or going to) the battery is accounted as work. Under this assumptions,
the amount of work we extract during one cycle is given by the changing in the average energy
of the system, that is
\begin{equation} \label{work_definition}
\Delta W = \Tr{P}{H_P \left( \rho_P - \tilde{\rho}_P  \right)},
\end{equation}
where $\rho_P$ is the initial passive state, and $\tilde{\rho}_P$ is the final state, whose probability
distribution is $\left\{ \p'_0 , \p'_1 , \p'_2 \right\}$. We can express the amount
of extracted work in terms of the energy gaps of the Hamiltonian $H_P$, as
\begin{equation} \label{work_qutrit}
\Delta W = \Delta E_{10} \left( \p'_0 - \p_0 \right) - \Delta E_{21} \left( \p'_2 - \p_2 \right),
\end{equation}
where this expression has been obtained by applying the normalisation constraint to the initial and final state
of the system.
\par
If we replace the probability distribution of the final state of the system, Eq.~\eqref{final_state_system},
into the expression of extracted work, Eq.~\eqref{work_qutrit}, we obtain that
\begin{equation} \label{work_cycle}
\Delta W =
\Delta E_{10}
\left( \sum_{j=1}^{m-1} \left( \p_1 \q_{j-1} - \p_0 \q_j \right) + \left( \p_1 \q_{m-1} - \p_0 \q_{m+n-1} \right) \right)
- \Delta E_{21}
\left( \sum_{j=m+1}^{m+n-1} \left( \p_1 \q_j - \p_2 \q_{j-1} \right) + \left( \p_1 \q_m - \p_2 \q_0 \right) \right).
\end{equation}
This expression can be highly simplified if we use the properties of the probability distribution of the
machine, Eqs.~\eqref{prob_dist_machine}.
In particular, from Eq.~\eqref{prob_dist_2} we find that
\begin{equation}
\p_1 \q_{j-1} - \p_0 \q_j = \p_1 \q_{j} - \p_0 \q_{j+1} \quad ; \quad \forall \, j = 1 , \ldots , m-2,
\end{equation}
while from \eqref{prob_dist_3} we have that
\begin{equation}
\p_1 \q_{m-2} - \p_0 \q_{m-1} = \p_1 \q_{m-1} - \p_0 \q_{m+n-1}.
\end{equation}
Together, these equations reduce the first bracket of Eq.~\eqref{work_cycle} into a single term,
\begin{equation}
\sum_{j=1}^{m-1} \left( \p_1 \q_{j-1} - \p_0 \q_j \right) + \left( \p_1 \q_{m-1} - \p_0 \q_{m+n-1} \right)
=
m \left( \p_1 \q_{m-1} - \p_0 \q_{m+n-1} \right).
\end{equation}
If we consider Eq.~\eqref{prob_dist_5}, instead, we find that
\begin{equation}
\p_1 \q_j - \p_2 \q_{j-1} = \p_1 \q_{j+1} - \p_2 \q_j  \quad ; \quad \forall \, j = m + 1 , \ldots , m + n - 2,
\end{equation}
while Eq.~\eqref{prob_dist_4} implies that
\begin{equation}
\p_1 \q_{m+1} - \p_2 \q_m = \p_1 \q_m - \p_2 \q_0.
\end{equation}
These two equations simplify the second bracket of Eq.~\eqref{work_cycle}, 
\begin{equation}
\sum_{j=m+1}^{m+n-1} \left( \p_1 \q_j - \p_2 \q_{j-1} \right) + \left( \p_1 \q_m - \p_2 \q_0 \right)
=
n \left( \p_1 \q_{m+n-1} - \p_2 \q_{m+n-2} \right).
\end{equation}
We can now use Eq.~\eqref{prob_dist_6} to show that
\begin{equation}
\p_1 \q_{m-1} - \p_0 \q_{m+n-1} = \p_1 \q_{m+n-1} - \p_2 \q_{m+n-2},
\end{equation}
which allows us to express the work we extract as
\begin{equation} \label{simply_work}
\Delta W = \left( m \, \Delta E_{10} - n \, \Delta E_{21} \right) \left( \p_1 \q_{m+n-1} - \p_2 \q_{m+n-2} \right).
\end{equation}
From the above equation we notice that the work extracted is factorised into an
Hamiltonian contribution and another contribution associated with the probability
distribution of the passive state. Then, for a given Hamiltonian $H_P$ such that $m
\, \Delta E_{10} > n \, \Delta E_{21}$, we will find that certain passive states allow for
work extraction (the ones in which $\p_1 \q_{m+n-1} > \p_2 \q_{m+n-2}$), while others do not. Therefore,
for every given Hamiltonian (that is, every $\Delta E_{10}$ and $\Delta E_{21}$) and for every given cycle
(that is, every $n$ and $m$), we find that the set of passive states is divided into two subsets, the ones which
allow for work extraction (we can call them \emph{activable states}), and the ones which do not.
In the following we will express the probability distribution
of $\rho_M$ in terms of the probability distribution of the passive state, so as to define these two subsets for each
Hamiltonian and cycle.
\par
As a first step, we want to express the first $m - 2$ elements of the sequence $\left\{ q_j \right\}_{j=0}^{m-1}$
in terms of last two elements, $\q_{m-2}$ and $\q_{m-1}$. Moreover, we express the first $n - 2$ elements of
$\left\{ q_j \right\}_{j=m}^{m+n-1}$ in terms of $\q_{m+n-2}$ and $\q_{m+n-1}$. This can be done by utilising the
equalities given in Eqs.~\eqref{prob_dist_2} and \eqref{prob_dist_5}, which we recast in the following way.
\begin{subequations}
\begin{align}
\label{sequence_1}
\q_j &= \left( 1 + \frac{\p_0}{\p_1} \right) \q_{j+1} - \frac{\p_0}{\p_1} \, \q_{j+2} \quad ; \quad \forall \, j = 0 , \ldots , m-3, \\
\label{sequence_2}
\q_j &= \left( 1 + \frac{\p_1}{\p_2} \right) \q_{j+1} - \frac{\p_1}{\p_2} \, \q_{j+2} \quad ; \quad \forall \, j = m , \ldots , m+n-3.
\end{align}
\end{subequations}
It can be proved (see the technical result~\ref{sequence_solution}) that the elements of the sequences can be
expressed as
\begin{subequations}
\begin{align}
q_j &= \T_1 \left( m - ( j + 2 ) \right) \q_{m - 2}
- \frac{\p_0}{\p_1} \, \T_1 \left( m - ( j + 3 ) \right) \q_{m - 1}
\quad ; \quad \forall \, j = 0 , \ldots , m-3, \\
q_j &= \T_2 \left( m + n - ( j + 2 ) \right) \q_{m + n - 2}
- \frac{\p_1}{\p_2} \, \T_2 \left( m + n - ( j + 3 ) \right) \q_{m + n - 1}
\quad ; \quad \forall \, j = m , \ldots , m+n-3,
\end{align}
\end{subequations}
where $\T_1 (h) = \sum_{l = 0}^{h} \left( \frac{\p_0}{\p_1} \right)^l$ and $\T_2 (h) = \sum_{l = 0}^{h} \left( \frac{\p_1}{\p_2} \right)^l$.
\par
We can now express, using Eqs.~\eqref{prob_dist_3} and \eqref{prob_dist_6}, the elements $\q_{m-2}$ and
$\q_{m-1}$ in terms of $\q_{m+n-2}$ and $\q_{m+n-2}$. From Eq.~\eqref{prob_dist_3} we obtain that
\begin{equation}
\q_{m-2} = \T_1(2) \, \q_{m+n-1} - \frac{\p_2}{\p_1} \, \T_1(1) \, \q_{m+n-2}.
\end{equation}
From Eq.~\eqref{prob_dist_6}, instead, we get that
\begin{equation}
\q_{m-1} = \T_1(1) \, \q_{m+n-1} - \frac{\p_2}{\p_1} \, \T_1(0) \, \q_{m+n-2}.
\end{equation}
Then, we can finally express $\q_{m+n-2}$ in terms of $\q_{m+n-1}$ through Eq.~\eqref{prob_dist_4}, and we obtain
\begin{equation}
\q_{m+n-2} = D(m,n) \, \q_{m+n-1},
\end{equation}
where the coefficient $D(m,n)$ is defined as
\begin{equation}
\label{D_coeff}
D(m,n) = \frac{\p_1}{\p_2} \frac{\T_1(m) + \frac{\p_1}{\p_2} \T_2(n-2)}{\T_1(m-1) + \frac{\p_1}{\p_2} \T_2(n-1)}.
\end{equation}
\par
Thanks to the above result, we can express the overall probability distribution of $\rho_M$ in terms of the
occupation probability of the state $\ket{m+n-1}_M$. Thus, we have that
\begin{subequations}
\label{machine_prob}
\begin{align}
\label{machine_prob_1}
\q_j &= \left( \T_1(m-j) - \frac{\p_2}{\p_1} \, D(m,n) \, \T_1 \left( m - ( j + 1 ) \right) \right) \q_{m+n-1}
\quad ; \quad j = 0, \ldots , m-1, \\
\label{machine_prob_2}
\q_j &= \left( \T_2 \left( m + n - ( j + 2) \right) \, D(m,n) - \frac{\p_1}{\p_2} \, \T_2 \left( m + n - ( j + 3 ) \right) \right) \q_{m+n-1}
\quad ; \quad j = m, \ldots , m+n-3, \\
\label{machine_prob_3}
\q_{m+n-2} &= D(m,n) \,\q_{m+n-1},
\end{align}
\end{subequations}
where it is possible to show that each $\q_j$, with $j = 0 , \ldots , m+n-2$, is positive if $\q_{m+n-1}$
is positive (see the technical result~\ref{positive_probabilities}). From the normalisation condition it then
follows that the sequence $\left\{ q_j \right\}_{j=0}^{m+n-1}$ is a proper probability distribution. Moreover,
the normalisation condition allows us to evaluate $\q_{m+n-1}$ as a function of the probability
distribution of the passive state $\rho_P$,
\begin{equation} \label{final_machine_prob}
\q_{m+n-1} = \frac{\T_1(m-1) + \frac{\p_1}{\p_2} \T_2(n-1)}
{\left( \T_1(m) + \frac{\p_1}{\p_2} \T_2(n-2) \right)^2 +
\left( \left( \frac{\p_1}{\p_2} \right)^n - \left( \frac{\p_0}{\p_1} \right)^m \right)
\left( \sum_{j=0}^m \T_1(j) - \frac{\p_1}{\p_2} \sum_{j=0}^{n-3} \T_2(j)  \right)}.
\end{equation}
From Eq.~\eqref{final_machine_prob} we can express all the other elements of $\left\{ q_j \right\}_{j=0}^{m+n-1}$
in terms of the probability distribution of $\rho_P$.
\par
We can now further characterise the amount of work extracted during our cycle. In fact, if we apply
Eq.~\eqref{machine_prob_3} into Eq.~\eqref{simply_work}, we obtain
\begin{equation}
\label{final_work}
\Delta W = \left( m \, \Delta E_{10} - n \, \Delta E_{21} \right)
\frac{ \p_1 \left( \left( \frac{\p_1}{\p_2} \right)^n - \left( \frac{\p_0}{\p_1} \right)^m \right)}
{\T_1(m-1) + \frac{\p_1}{\p_2} \T_2(n-1)} \, \q_{m+n-1},
\end{equation}
where the sign of $\Delta W$ depends on the sole terms $\left( m \, \Delta E_{10} - n \, \Delta E_{21} \right)$
and $\left( \left( \frac{\p_1}{\p_2} \right)^n - \left( \frac{\p_0}{\p_1} \right)^m \right)$, since the other factors are
always positive. Thus, for each cycle, we can characterise which passive states can be activated by that cycle,
that is, which states allow for work extraction during the cycle. The subset of activable states is
\begin{align} \label{activation_set}
R^{+}_{m,n} = \bigg\{ \rho_P \ \text{passive} \ \bigg| 
\ &\left( \frac{\p_1}{\p_2} \right)^n > \left( \frac{\p_0}{\p_1} \right)^m
\text{when} \
m \, \Delta E_{10} - n \, \Delta E_{21} > 0 \nonumber \\
\vee
\ &\left( \frac{\p_1}{\p_2} \right)^n < \left( \frac{\p_0}{\p_1} \right)^m
\text{when} \
m \, \Delta E_{10} - n \, \Delta E_{21} < 0
\bigg\},
\end{align}
where this region clearly depends on the Hamiltonian of the system $H_P$, and on the number of
swaps performed during the cycle, $m$ and $n$.
\subsection{The final state of the system}
\label{final_state_system_sec}
Let us consider the final state of the passive system after we have applied the cycle $S_{m,n}$. In
Eq.~\eqref{final_state_system} we have shown the probability distribution of $\tilde{\rho}_P$ as a
function of $\left\{ \q_i \right\}_{i=0}^{m+n-1}$. Thanks to the constraints introduced in
Eqs.~\eqref{prob_dist_machine}, we can simplify the form of $\tilde{\rho}_P$, so that we obtain
\begin{subequations}
\label{sm_final_prob_dist}
\begin{align}
\p'_0 &= \p_0 + m \, \Delta \mathrm{P}, \\
\p'_1 &= \p_1 - \left( m + n \right) \, \Delta \mathrm{P}, \\
\p'_2 &= \p_2 + n \, \Delta \mathrm{P}.
\end{align}
\end{subequations}
We can
easily notice that the cycle acts on the passive state by modifying the original probabilities by multiples of
\begin{equation}
\label{sm_prob_unit}
\Delta \mathrm{P} = \frac{ \p_1 \, \q_{m+n-1} }
{\T_1(m-1) + \frac{\p_1}{\p_2} \T_2(n-1)} \,
\left( \left( \frac{\p_1}{\p_2} \right)^n - \left( \frac{\p_0}{\p_1} \right)^m \right).
\end{equation}
The expression of the final state $\tilde{\rho}_P$ allows us to understand how the cycle operates over
the system when work is extracted. In particular, we can consider the evolution of the system in two
different situations, linked to the two possible scenarios of Eq.~\eqref{activation_set}.
\par
\begin{figure}[!ht]
\center
\includegraphics[width=0.8\textwidth]{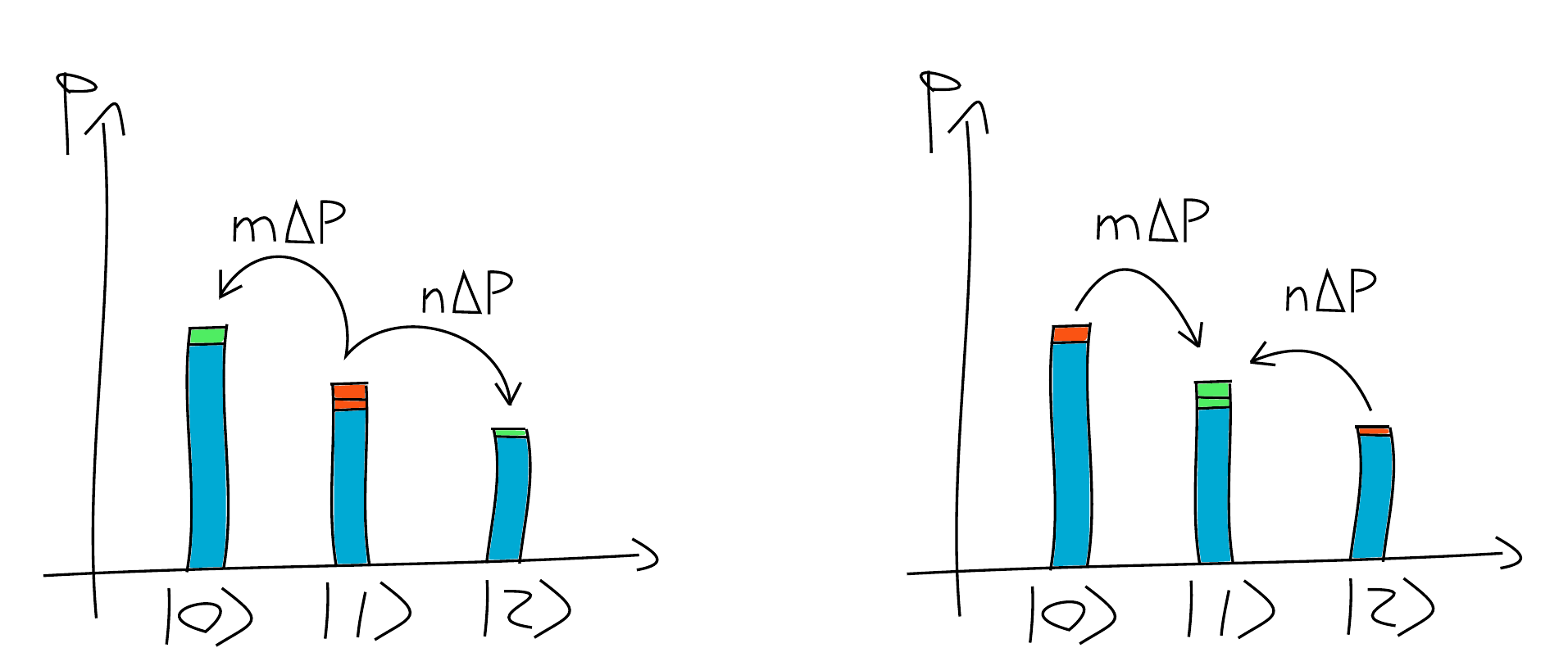}
\caption{(Left) The action of the cycle on the system, when work is extracted and the Hamiltonian
$H_P$ is such that $m \, \Delta E_{10} > n \, \Delta E_{21}$. The probability of occupation of $\ket{1}_P$
is reduced, while the other two probabilities of occupation increase. (Right) The action of the cycle on
the system, when work is extracted and the Hamiltonian $H_P$ is such that $m \, \Delta E_{10} < n \,
\Delta E_{21}$. The map acts on the system in the opposite way compared to the previous scenario.}
\label{fig2:map_evolution}
\end{figure}

Suppose that $H_P$ is such that $m \, \Delta E_{10} > n \, \Delta E_{21}$. Then, from the conditions
in $R^{+}_{m,n}$, we can verify that $\Delta \mathrm{P} > 0$, so that the map is depleting the population
of the state $\ket{1}_P$, while increasing the populations of both $\ket{0}_P$ and $\ket{2}_P$ (see
Fig.~\ref{fig2:map_evolution}, left plot). Work is extracted from the cycle since the energy gained while
moving $m \, \Delta \mathrm{P}$ from $\p_1$ to $\p_0$ is bigger than the energy paid to move
$n \, \Delta \mathrm{P}$ from $\p_1$ to $\p_2$. In Sec.~\ref{features_activation_map}, we show that the entropy
of the system has to increase during the transformation. This is achieved since $\p_1$ gets closer to
$\p_2$ after the cycle.
\par
Let us consider the case in which $H_P$ is such that $m \, \Delta E_{10} < n \, \Delta E_{21}$. Then, from the conditions
in $R^{+}_{m,n}$, we can verify that $\Delta \mathrm{P} < 0$, so that the map is depleting the populations
of the states $\ket{0}_P$ and $\ket{2}_P$, while increasing the populations of $\ket{1}_P$ (see
Fig.~\ref{fig2:map_evolution}, right plot). Work is extracted from the cycle since the energy gained while
moving $n \, \Delta \mathrm{P}$ from $\p_2$ to $\p_1$ is bigger than the energy paid to move
$m \, \Delta \mathrm{P}$ from $\p_0$ to $\p_1$. Moreover, the entropy of the system increases since
$\p_0$ gets closer to $\p_1$ after the cycle.
\par
It is worth noting that the final state $\tilde{\rho}_P$ can be active. This happen, in the case of
$m \, \Delta E_{10} > n \, \Delta E_{21}$, when $\p'_1 < \p'_2$. In the other case, we obtain a
final active state if $\p'_0 < \p'_1$. In these situations, not only are we able to extract work from
the passive state $\rho_P$ during the cycle, but we can also perform a local unitary operation
(permuting $\ket{1}_P$ and $\ket{2}_P$ in the first case, and  $\ket{0}_P$ and $\ket{1}_P$ in
the second) which allows for additional work extraction. It is also possible for the final state of
the system to be passive, and to still lie inside the activable region $R^{+}_{m,n}$. Due to the
correlation created between system and machine, however, this state cannot be used
again, at least not with the same machine.
\section{Asymptotic behaviour of the machine}
\label{Asympt_behaviour}
We are now interested in the study of the cycle $S_{m,n}$ when the size of the machine
(as well as the number of hot and cold swaps) tends to infinity. In particular, we are interested in
the form of the probability distribution of the machine, the work extracted, and the final
state of the passive system. Let us consider the Hamiltonian $H_P$ of the main (qutrit) system.
We know that, for any Hamiltonian $H_P$, there exists two integer numbers $N$ and $M$ such
that $M \, \Delta E_{10} = N \, \Delta E_{21}$. We now consider a passive state $\rho_P$ describing
this system whose probability distribution satisfies $\left( \p_1/\p_2 \right)^N > \left( \p_0/\p_1 \right)^M$.
Notice that this condition implies that the state is in the subset of passive states denoted by $R_1$
(see the main text, Fig.~5), or equivalently it implies that the hot virtual temperature is associated with
the pair of states $\ket{0}_P$ and $\ket{1}_P$. One could analyse the opposite situation as well, but
the results we obtain would be analogous, due to the symmetry of the problem with respect to the
hot and cold interactions.
\par
In order to perform the asymptotic expansion of the probability distribution of the machine,
Eqs.~\eqref{machine_prob}, we first want to define how the ratio $\frac{n}{m}$ behaves as the
number of hot and cold swaps goes to infinity. We set this fraction equal to $\alpha$,
so that $n = \alpha \, m$, and we define a range for this parameter, due to the constraints we
set on the passive state. Indeed, if we want to extract work, we need $m$ and $n$ to satisfy one
of the two conditions in Eq.~\eqref{activation_set}, and in particular, since we assume the passive
state to be in the region $R_1$, we need $m \, \Delta E_{10} > n \, \Delta E_{21}$ and
$\left( \p_1/\p_2 \right)^n > \left( \p_0/\p_1 \right)^m$. The two inequalities implies that
\begin{equation}
\label{range_alpha}
\frac{ \log \frac{\p_0}{\p_1} }{ \log \frac{\p_1}{\p_2} } < \alpha < \frac{\Delta E_{10}}{\Delta E_{21}},
\end{equation}
where it is easy to verify that the lower bound is smaller than the upper one,
due to the fact that $\rho_P \in R_1$.
\par
We can now use the assumptions made on the cycle (that is, on the parameters $m$ and $n$)
and on the initial passive state in order to expand the probability distribution of $\rho_M$ for
$m, n \rightarrow \infty$. As a first step, let us consider the coefficient $D(m,n)$ presented in
Eq.~\eqref{D_coeff}. When $m$ and $n$ tends to infinity, we find that
\begin{equation}
D(m,n) \approx 1 + \left( \frac{\p_0}{\p_1} \right)^m \left( \frac{\p_2}{\p_1} \right)^n
\frac{ \left( \p_0 - \p_2 \right) \left( \p_1 - \p_2 \right) }{ \left( \p_0 - \p_1 \right) }
+ O \left( \left( \frac{\p_2}{\p_1} \right)^n ; 
\left( \frac{\p_0}{\p_1} \right)^{2 m} \left( \frac{\p_2}{\p_1} \right)^{2 n} \right),
\end{equation}
where it is easy to verify that the term $\left( \p_0/\p_1 \right)^m \left( \p_2/\p_1 \right)^n
\rightarrow 0$ as $m,n \rightarrow \infty$, and that both $\left( \p_2/\p_1 \right)^n$ and
$\left( \p_0/\p_1 \right)^{2 m} \left( \p_2/\p_1 \right)^{2 n}$ tends to 0 faster that this first
term. However, we cannot say which one is the fastest without further assumptions, and
that is the reason we keep both in the $O$.
\par
Once the expansion of $D(m,n)$ is known, we can focus on the probability distribution of
the machine. For simplicity, we consider the distribution in Eqs.~\eqref{machine_prob}, where
$\q_{m+n-1}$ is not defined yet; we will define it through the normalisation condition once the
asymptotic expansion has been performed. We find that
\begin{subequations}
\begin{align}
\q_j &\approx \q_{m+n-1} \left( \frac{\p_0 - \p_2}{\p_0 - \p_1}
+ O \left( \left( \frac{\p_0}{\p_1} \right)^m \left( \frac{\p_2}{\p_1} \right)^n \right) \right)
\left( \frac{\p_0}{\p_1} \right)^{m-j} \quad ; \quad j = 0, \ldots, m-1,\\
\q_j &\approx \q_{m+n-1} \left( \frac{\p_2}{\p_1} \frac{\p_0 - \p_2}{\p_0 - \p_1}
+ O \left( \left( \frac{\p_0}{\p_1} \right)^m \left( \frac{\p_2}{\p_1} \right)^n \right) \right)
\left( \frac{\p_0}{\p_1} \right)^m \left( \frac{\p_1}{\p_2} \right)^{m-j} \quad ; \quad j = m, \ldots, m+n-3,\\
\q_{m+n-2} &\approx \q_{m+n-1}
\left( 1 + O \left( \left( \frac{\p_0}{\p_1} \right)^m \left( \frac{\p_2}{\p_1} \right)^n \right) \right).
\end{align}
\end{subequations}
We are now able to obtain the value of $\q_{m+n-1}$ by imposing the normalisation condition over
the asymptotic probability distribution of the machine. We find that
\begin{equation}
\q_{m+n-1} \approx 
\left(
\frac{ \left( \p_1 - \p_2 \right) \left( \p_0 - \p_1 \right)^2 }{ \p_1 \left( \p_0 - \p_2 \right)^2 }
+ O \left( \left( \frac{\p_0}{\p_1} \right)^m \left( \frac{\p_2}{\p_1} \right)^n \right)
\right)
\left( \frac{\p_1}{\p_0} \right)^m,
\end{equation}
that is, $\q_{m+n-1}$ tends to $0$ as $\left( \p_1/\p_0 \right)^m$ for $m \rightarrow \infty$.
Notice that the same result can be obtained by expanding Eq.~\eqref{final_machine_prob}.
If we send $m$ and $n$ to infinity, we find that the asymptotic probability distribution of the
machine is
\begin{subequations}
\begin{align}
\q_j &\approx \frac{\left( \p_1 - \p_2 \right) \left(  \p_0 - \p_1  \right)}{\p_1 \left( \p_0 - \p_2 \right)}
\left( \frac{\p_0}{\p_1} \right)^{-j} \quad ; \quad j = 0, \ldots, m-1,\\
\q_j &\approx \frac{\p_2 \left( \p_1 - \p_2 \right) \left(  \p_0 - \p_1  \right)}{\p_1^2 \left( \p_0 - \p_2 \right)}
\left( \frac{\p_1}{\p_2} \right)^{m-j} \quad ; \quad j = m, \ldots, m+n-3,\\
\q_{m+n-2} \approx \q_{m+n-1} &\approx
\frac{ \left( \p_1 - \p_2 \right) \left( \p_0 - \p_1 \right)^2 }{ \p_1 \left( \p_0 - \p_2 \right)^2 }
\left( \frac{\p_1}{\p_0} \right)^m.
\end{align}
\end{subequations}
\par
We can now investigate how the probability distribution of the main system changes, and evaluate
 the asymptotic work extracted $\Delta W$ during on cycle. Let us consider the probability
unit $\Delta \mathrm{P}$, introduced in Eq.~\eqref{sm_prob_unit}. If we set $m$ and $n$ to infinity,
we have that
\begin{equation}
\Delta \mathrm{P} \approx
\frac{ \left( \p_1 - \p_2 \right)^2 \left( \p_0 - \p_1 \right)^2 }{ \p_1 \left( \p_0 - \p_2 \right)^2 }
\left( \frac{\p_1}{\p_0} \right)^m,
\end{equation}
that tends to 0 with an exponential scaling. Therefore, the heat engine with infinite-dimensional thermal
machine only modifies the passive states by an infinitesimal amount. As a consequence, the
work extracted has to be infinitesimal as well. Indeed, by considering Eq.~\eqref{final_work}
it is easy to show that $\Delta W$ tends to $0$ as $m, n \rightarrow \infty$, since $\Delta W$ is
proportional to $\Delta \mathrm{P}$ (modulo a multiplying factor proportional to $m$, which
tends to infinity more slowly than $\left( \p_1/\p_0 \right)^m$ tends to $0$).
\subsection{Final state and work extraction over multiple cycles}
In the previous section we have seen that, when the machine is infinitely large,
we only modify the passive state infinitesimally. We can then consider the situation in
which we are given an infinite number of these machines, and we want to evolve
the passive state (and extract work) by sequentially applying our cycle with the help of
these machines. In order to study the evolution of the passive state, we can consider its
probability distribution after one cycle, see Eqs.~\eqref{sm_final_prob_dist}. These equations
can be recast as differential equations, since $\Delta \mathrm{P} \rightarrow 0$ in this
scenario. It is easy to verify that the differential equations which govern the evolution
of the passive state are
\begin{subequations}
\label{diff_eqs}
\begin{align}
\frac{ \rmd \p_0 }{ \rmd t } &=
\frac{ \left( \p_1 - \p_2 \right)^2 \left( \p_0 - \p_1 \right)^2 }{ \p_1 \left( \p_0 - \p_2 \right)^2 }, \\
\frac{ \rmd \p_1 }{ \rmd t } &= - \left( 1 + \alpha \right)
\frac{ \left( \p_1 - \p_2 \right)^2 \left( \p_0 - \p_1 \right)^2 }{ \p_1 \left( \p_0 - \p_2 \right)^2 },
\end{align}
\end{subequations}
where $\alpha = \frac{n}{m}$ takes values in the range given by Eq.~\eqref{range_alpha}, and we define
\begin{equation}
\frac{ \rmd \p_i }{ \rmd t } = \lim_{m \rightarrow \infty} \frac{\p'_i - \p_i}{ \Delta \p ( m )}
\quad , \quad
\Delta \p ( m ) = m \left( \frac{\p_1}{\p_0} \right)^m
\quad
\text{for} \ i = 0, 1,
\end{equation}
with $\left\{ \p_0, \, \p_1, 1 - \p_0 - \p_1 \right\}$ the probability distribution of the state before the cycle, and
$\left\{ \p'_0, \, \p'_1, 1 - \p'_0 - \p'_1 \right\}$ the distribution of the state after the cycle. The continuous
parameter $t$ is here related to the number of cycles we perform on the system. It is worth noting that
Eqs.~\ref{diff_eqs} share a common (positive) factor. Therefore we have that, as time goes on, the probability
of occupation of $\ket{0}_P$ increases, while the one of $\ket{1}_P$ decreases (as expected from the discussion
in Sec.~\ref{final_state_system_sec}). Moreover, since $\alpha > 0$, the increase in the former is slower than the
decreasing of the latter.
\par
The two differential equations can be reshaped in a single, more helpful one,
\begin{equation}
\label{main_diff_eq}
\frac{ \rmd \p_1 }{ \rmd t } = - \left( 1 + \alpha \right) \frac{ \rmd \p_0 }{ \rmd t },
\end{equation}
and we can investigate the solution of this equation for $\alpha$ close to its limiting values.
As a first step, let us consider the case in which $\alpha = \frac{\Delta E_{10}}{\Delta E_{12}} -
\frac{1}{m} \approx \frac{\Delta E_{10}}{\Delta E_{12}}$. Then, the solution of Eq.~\eqref{main_diff_eq}
is
\begin{equation}
\label{first_sol}
\p_1(t) = - \left( 1 + \frac{\Delta E_{10}}{\Delta E_{12}} \right) \big( \p_0(t) - \p_0(t=0) \big) + \p_1(t=0),
\end{equation}
where $\left\{ \p_0(t), \, \p_1(t), 1 - \p_0(t) - \p_1(t) \right\}$ is the probability distribution of the state
of the system at time $t$, and $t=0$ is the initial time (when the system is in $\rho_P$).
If we rearrange Eq.~\eqref{first_sol}, we see that it is
equivalent to the following constraint for the evolved state
\begin{equation}
\tr{H_P \, \rho_P(t)} = \tr{H_P \, \rho_P} \quad \forall \ t \geq 0,
\end{equation}
that is, the evolution conserves the energy of the system (equivalently, no work is extracted during the
evolution). It is easy to see, for instance by representing the solution of Eq.~\eqref{first_sol} in a two-dimensional
plot of $\p_1$ versus $\p_0$, that the passive state is moving toward the set of thermal states, that are
the steady states of this evolution. In fact, when a thermal state is considered, we find that $\left( \p_1/
\p_2 \right)^{\alpha} = \left( \p_0/\p_1 \right)$, which implies $\Delta \mathrm{P} = 0$.
Thus, after enough time $t$ is passed, we find that the initial passive state $\rho_P$ has been mapped
into the thermal state with inverse temperature $\beta_{\text{min}}$, where
\begin{equation}
\beta_{\text{min}} \ : \ \tr{H_P \, \tau_{\beta_{\text{min}}}} = \tr{H_P \, \rho_P}
\quad , \quad
\tau_{\beta_{\text{min}}} = \frac{e^{-\beta_{\text{min}} H_P}}{Z_{\text{min}}},
\end{equation}
and $Z_{\text{min}}$ is the partition function of the system at temperature $\beta_{\text{min}}^{-1}$.
\par
We can now consider the case in which $\alpha = \frac{ \log \p_0 - \log \p_1 }{ \log \p_1 - \log \p_2 } +
\frac{1}{m} \approx \frac{ \log \p_0 - \log \p_1 }{ \log \p_1 - \log \p_2 }$, that is, when its value is close
to its lower bound. We notice that, in this case, $\alpha$ itself depends on the probability distribution
of the passive state. Then, if we replace $\alpha$ with its lower bound in Eq.~\eqref{main_diff_eq} we
obtain
\begin{equation}
\log \p_0 \, \frac{ \rmd \p_0 }{ \rmd t } + \log \p_1 \, \frac{ \rmd \p_1 }{ \rmd t } + \log \p_2 \, \frac{ \rmd \p_2 }{ \rmd t } = 0,
\end{equation}
which, if integrated between time $0$ and time $t$, gives the following constraint on the entropy of the
evolved states
\begin{equation}
S \left( \rho_P(t) \right) = S \left( \rho_P \right) \quad \forall \ t \geq 0,
\end{equation}
where $S(\rho) = - \tr{\rho \log \rho}$ is the Von Neumann entropy. Therefore, the evolution of the passive
state has to preserve the entropy of the system, and the state is moving toward the set of thermal states.
For $t \rightarrow \infty$, the system is in the thermal state with inverse temperature $\beta_{\text{max}}$,
where
\begin{equation}
\beta_{\text{max}} \ : \ S \left(  \tau_{\beta_{\text{max}}} \right) = S \left( \rho_P \right)
\quad , \quad
\tau_{\beta_{\text{max}}} = \frac{e^{-\beta_{\text{max}} H_P}}{Z_{\text{max}}},
\end{equation}
and $Z_{\text{max}}$ is the partition function of the system at temperature $\beta_{\text{max}}^{-1}$.
\par
Thus, when we set $\alpha$ equal to its limiting values, the evolution of the passive state can either follow
a trajectory in which energy is conserved, or in which entropy is conserved. However, all intermediate
trajectories can be achieved by imposing a different $\alpha$ inside the range specified by
Eq.~\eqref{range_alpha}, and consequently all passive states with lower or equal energy,
and greater or equal entropy that $\rho_P$ can be reached.
\section{Activation maps}
\label{features_activation_map}
Consider a specific family of CPT maps which allow for work extraction from a system described by a passive
state. The maps of this family, which we call \emph{activation maps}, can be represented by unitary operations
acting globally on both the main system and an ancilla, such that the local state of the ancillary system is preserved.
The cycle of Sec.~\ref{sec_general_cycle} is a particular instance of these activation maps, and in the following we
study the main properties of this family. Let us consider a system $S$ with Hamiltonian
$H_S$, described by the state $\rho_S$ (this state does not need to be passive). The energy that
we extract from the system when we evolve it with the unitary operator $U_S$ is given by the difference
in average energy between the initial and final state,
\begin{equation} \label{work_system}
\Delta W_S = \Tr{S}{H_S \left( \rho_S - U_S \, \rho_S \, U_S^{\dagger} \right)}.
\end{equation}
We assume this energy to be stored in an implicit battery, and we refer to it as work.
If the state is passive, then $\Delta W_S \leq 0$, that is, we cannot extract work. If the
state is active, we can find some unitary operations that allow for a positive work extraction.
In particular, the maximum work we can extract is
\begin{equation} \label{max_work}
\Delta W_S^{\text{max}} = \Tr{S}{H_S \left( \rho_S - \rho_S^{\text{pass}} \right)},
\end{equation}
where the state $\rho_S^{\text{pass}}$ is the passive state obtained from the initial state $\rho_S$.
In the literature, $\Delta W_S^{\text{max}}$ is known as \emph{ergotropy}, see Ref.~\cite{allahverdyan_maximal_2004}.
This quantity is $0$ if the initial state is passive, and positive otherwise.
\par
We now add an ancillary system $A$ with a trivial Hamiltonian, described by the state $\sigma_A$, and
we consider the family of maps
\begin{equation} \label{activation_map}
\Lambda \left( \rho_S \right) = \Tr{A}{U_{SA} \left( \rho_S \otimes \sigma_A \right) U_{SA}^{\dagger}},
\end{equation}
where the unitary operator $U_{SA}$ acts globally over system and ancilla, and we require that the
final local state of the ancilla is equal to the initial one, that is,
\begin{equation} \label{act_map_cond}
\sigma_A = \Tr{S}{U_{SA} \left( \rho_S \otimes \sigma_A \right) U_{SA}^{\dagger}}.
\end{equation}
Notice that the global evolution can create correlations between system and ancilla, and
our sole constraint regards the local state of the ancilla. The work extracted during the evolution
is given by
\begin{equation} \label{work_system_catalyst}
\Delta W_{SA} = \Tr{S}{H_{S} \left( \rho_S - \Lambda( \rho_S ) \right)},
\end{equation}
where the only contribution is given by the energy difference in the system, due to the absence
of any interaction term between system and ancilla, and to the fact that the final state of the ancilla
is equal to its initial one.
\par
We can now introduce the notion of \emph{activation} of a quantum state,
\begin{definition} \label{activation}
Let us consider a system $S$ with Hamiltonian $H_S$, described by the state $\rho_S$.
Then, we say that $\rho_S$ can be \emph{activated} iff there exists an ancillary system $A$
with trivial Hamiltonian, described by the state $\sigma_A$, and an activation map $\Lambda$
as in Eq.~\eqref{activation_map}, satisfying the condition of Eq.~\eqref{act_map_cond}, such that
\begin{equation}
\Delta W_{SA} > \Delta W^{\mathrm{max}}_S
\end{equation}
that is, if we can extract more work from $\rho_S$ by acting with $\Lambda$ than we can do by
acting with any unitary operation.
\end{definition}
As we noticed before, an example of activation map is the one used in our passive engine,
Eq.~\eqref{final_state_system}, where the ancillary system is the machine, and the
global unitary operation is $S_{m,n}$.
\subsection{General properties of the final state of an activation map}
Although the family of maps introduced in the previous section is extremely general, we can still use their
definition to derive some properties of the final state $\Lambda \left( \rho_S \right)$. The first, trivial property
consists in the fact that the final state of an activation map has to have a lower energy than the one
possessed by a the passified version of the initial state,
\begin{equation} \label{energy_constraint}
 \Tr{S}{H_S \, \rho_S^{\text{pass}}} > \Tr{S}{H_S \, \Lambda \left( \rho_S \right)},
\end{equation}
where this condition is obtained by replacing Eqs.~\eqref{max_work} and \eqref{work_system_catalyst}
into Def.~\ref{activation}.
\par
A second property regards the entropy of the final state. Due to the invariance of Von Neumann
entropy under unitary operations, its sub-additivity, and the constraint on the local state of the
machine, Eq.~\eqref{act_map_cond}, we can show that
\begin{equation} \label{entropy_constraint}
S(\rho_S) \leq S(\Lambda \left( \rho_S \right)),
\end{equation}
that is, the entropy of the system cannot decrease during the evolution through $\Lambda$, and it
increases if correlations create between system and machine.
\par
If we use the two constraints on $\Lambda \left( \rho_S \right)$ together, we can show that any
completely passive state cannot be activated. In this case, in fact, we have that $\rho_S =
\rho_S^{\text{pass}} = \tau_{\beta}$, that is, the state under examination is the thermal state of
Hamiltonian $H_S$ for a certain $\beta \in [0,\infty]$. But we know that this state is the one with
minimum energy for a given entropy, or, vice versa, the one with maximum entropy for given energy.
Then, we cannot find another state $\Lambda \left( \rho_S \right)$ such that the two conditions of
Eqs.~\eqref{energy_constraint} and \eqref{entropy_constraint} are satisfied at the same time. This
implies that any completely passive state cannot be activated, pure ground state and maximally-mixed
state included.
\par
We can also consider a generic pure state $\rho_S = \ket{\psi}\bra{\psi}$. The corresponding passified
state is the ground state $\ket{0}$. From Eq.~\eqref{energy_constraint} it follows
that the final state of $\Lambda$ has to have a lower energy than $\rho_S^{\text{pass}}$. But since the
passified state we obtain, $\ket{0}$, is by definition the state with minimum energy, we cannot satisfy this condition.
Thus, we cannot activate, in the sense of Def.~\ref{activation}, any pure state $\ket{\psi}$.
\subsection{Asymptotic work extraction from passive states}
\label{asympt_work_extr}
It was proved by Alicki {\it et al.} (Ref.~\cite{alicki_entanglement_2013}) that, when an infinite number
of copies of a passive state $\rho_S$ are considered, the optimal extractable work per single copy is
given by
\begin{equation}
\Delta W_{\text{opt}} = \Tr{S}{H_S \left( \rho_S - \tau_{\beta_{\text{max}}} \right)},
\end{equation}
where $\tau_{\beta_{\text{max}}}$ is the thermal state with inverse temperature $\beta_{\text{max}}$
such that $S(\tau_{\beta_{\text{max}}}) = S(\rho_S)$. We want to compare the work
extracted in the asymptotic limit with the work extracted with a generic activation map $\Lambda$.
This comparison can be easily carried out using the main properties of the final state $\Lambda
( \rho_S )$, see Eqs.~\eqref{energy_constraint} and \eqref{entropy_constraint}, together with the
properties of $\tau_{\beta_{\text{max}}}$.
\par
For any given final state of the system $\Lambda( \rho_S )$, there always exists an inverse
temperature $\hat{\beta}$, and a thermal state $\tau_{\hat{\beta}}$ at that temperature, such that
$S(\tau_{\hat{\beta}}) = S \left( \Lambda ( \rho_S ) \right)$.
Since the state $\tau_{\hat{\beta}}$ is thermal, we have that its energy is minimum, that is,
\begin{equation} \label{ineq_energy_1}
\Tr{S}{ H_S \, \Lambda ( \rho_S ) } \geq \Tr{S}{ H_S \tau_{\hat{\beta}} }.
\end{equation}
Moreover, from Eq.~\eqref{entropy_constraint} it follows that the entropy of $\tau_{\hat{\beta}}$ is
greater than the entropy of the state $\tau_{\beta_{\text{max}}}$, introduced in the previous paragraph.
By considering this entropic condition together with the free energy difference
$F_{\beta_{\text{max}}} ( \tau_{\hat{\beta}} ) - F_{\beta_{\text{max}}} ( \tau_{\beta_{\text{max}}} ) \geq 0$, we obtain
that the state $\tau_{\hat{\beta}}$ is more energetic than $\tau_{\beta_{\text{max}}}$, that is,
\begin{equation} \label{ineq_energy_2}
\Tr{S}{ H_S \tau_{\hat{\beta}} } \geq \Tr{S}{ H_S \tau_{\beta_{\text{max}}} }.
\end{equation}
From the above inequalities we have that
\begin{equation}
\Delta W_{\text{opt}} - \Delta W_{SA} = \Tr{S}{ H_S \, \Lambda ( \rho_S ) } - \Tr{S}{ H_S \tau_{\beta_{\text{max}}} }
\overset{\text{Eq.~\eqref{ineq_energy_1}}}{\geq}
\Tr{S}{ H_S \tau_{\hat{\beta}} } - \Tr{S}{ H_S \tau_{\beta_{\text{max}}} }
\overset{\text{Eq.~\eqref{ineq_energy_2}}}{\geq}
0.
\end{equation}
Therefore, the energy we extract with the aid of an activation map $\Lambda$ is always equal
or lower than the energy (per single copy) that we extract by acting over an infinite number of copies
of the passive state with a global unitary operator, $\Delta W_{\text{opt}} \geq \Delta W_{SA}$.
\par
Thus, $\Delta W_{\text{opt}}$ is an upper bound for the work extracted by any activation map
$\Lambda$. In Refs.~\cite{alicki_entanglement_2013, sparaciari_resource_2016} it was shown
that this upper bound can be actually achieved by acting over infinite
many copies of the system with a global unitary operation. In this paper, instead, we have shown
that the extraction of an amount of work equal to $\Delta W_{\text{opt}}$ is also achievable by
acting on a single copy of the state. However, one needs to utilise infinite many infinite-dimensional
machines to do so, as we showed in Sec.~\ref{Asympt_behaviour}.
\subsection{Ancilla as part of a bigger thermal bath}
Consider the case in which the ancilla utilised in $\Lambda$ is just a subsystem
of an infinite thermal reservoir at temperature $\beta^{-1}$. In this situation, we have to explicitly define an
Hamiltonian $H_A$ (where we have the freedom to rigidly translate the spectrum of this Hamiltonian), so
that the state of the ancilla $\sigma_A$ coincides with the thermal state $\tau_{\beta}^{(A)} = e^{-\beta H_A}/Z_A$.
\par
As we have seen, the map $\Lambda$ lowers the energy of the system and builds correlations between
system and ancilla, while preserving the local state of the ancillary system. If we consider the ancilla as part
of the infinite bath, then we see that $\Lambda$ extracts work from the passive state while no heat is
exchanged with the bath (as the local state of the ancilla is unchanged). In the following we show that the
energy extracted during this transformation is always lower than the difference in free energy between the
initial state $\rho_S$ and the thermal state $\tau_{\beta}^{(S)} = e^{-\beta H_S}/Z_S$. Even in the case in
which $\Lambda$ maps $\rho_S$ into $\tau_{\beta}^{(S)}$, the work extracted is not optimal, as part of this
work is locked inside the correlations between system and ancilla. In order to extract the remaining work from
the correlations, and thus to perform optimal work extraction, we have to exploit the infinite thermal reservoir,
exchanging an amount of heat proportional to the difference in entropy between $\tau_{\beta}^{(S)}$ and
$\rho_S$. It is worth noting that, although this second operation allows us to extract an higher amount of work
than the one obtained with the sole $\Lambda$, we do not consider it as an allowed operation in our framework,
as it requires an additional ancillary system (the bath) with infinite dimension.
\par
During the first operation we map the initial state $\rho_S$ into the final one $\Lambda ( \rho_S )$.
This final state might or might not be a thermal state of $H_S$, and the sole constraints we have are
given by Eqs.~\eqref{energy_constraint} and \eqref{entropy_constraint} (energy has to decrease while
entropy has to increase). The work we extract is the energy difference between the initial and final state,
as we show in Eq.~\eqref{work_system_catalyst},
\begin{equation}
\Delta W_1 =  \Tr{S}{H_{S} \left( \rho_S - \Lambda( \rho_S ) \right)},
\end{equation}
which is positive by definition, since we assume $\Lambda$ to be an activation map, see Def.~\ref{activation}.
The final state of system and ancilla is $\tilde{\rho}_{SA} = U_{SA} \left( \rho_S \otimes \tau_{\beta}^{(A)}
\right) U_{SA}^{\dagger}$, and correlations are present, quantified by the mutual information
\begin{equation}
I \left( \tilde{S} : \tilde{A} \right) = S(\Lambda( \rho_S )) + S(\tau_{\beta}^{(A)}) - S(\tilde{\rho}_{SA}).
\end{equation}
The heat $Q_1$ exchanged during this transformation is equal to $0$,
as the local state of the bath does not change.
\par
We now use the power of the infinite thermal reservoir to extract the last part of work from the state
$\tilde{\rho}_{SA}$, by mapping it into $\tau_{\beta}^{(S)} \otimes \tau_{\beta}^{(A)}$. In this case,
work is given by the free energy difference between the two states, that is
\begin{equation}
\Delta W_2 = F_{\beta} \left( \tilde{\rho}_{SA} \right) - F_{\beta} \left( \tau_{\beta}^{(S)} \otimes \tau_{\beta}^{(A)} \right)
= \frac{1}{\beta} \left( D \left( \Lambda( \rho_S ) || \tau_{\beta}^{(S)} \right) + I \left( \tilde{S} : \tilde{A} \right) \right),
\end{equation}
where $ D \left( \Lambda( \rho_S ) || \tau_{\beta}^{(S)} \right) = \beta \left( F_{\beta} \left( \Lambda( \rho_S ) \right) -
F_{\beta} \left( \tau_{\beta}^{(S)} \right) \right)$ is the relative entropy between $\Lambda( \rho_S )$ and
$ \tau_{\beta}^{(S)}$. Since both the relative entropy and the mutual information are non-negative quantities,
we have that work is indeed extracted during this second process. The heat exchanged in this second
transformation is equal to the entropy difference (modulo the multiplicative constant $\beta^{-1}$) between
the final and initial state
\begin{equation}
Q_2 = \frac{1}{\beta} \left( S \left( \tau_{\beta}^{(S)} \otimes \tau_{\beta}^{(A)} \right)
- S \left( \tilde{\rho}_{SA} \right) \right)
= \frac{1}{\beta} \left( S \left( \tau_{\beta}^{(S)} \right) - S \left( \rho_{S} \right) \right),
\end{equation}
where the last equality follows from the invariance under unitary operations of the Von Neumann entropy.
\par
If we now consider the two transformations as a single one, we see that the total work extracted is
\begin{equation}
\Delta W_{\text{tot}} = \Delta W_1 + \Delta W_2 = F_{\beta} \left( \rho_{S} \right) - F_{\beta} \left( \tau_{\beta}^{(S)} \right),
\end{equation}
that is, $\Delta W_{\text{tot}}$ is optimal, and the heat exchanged is $Q_2$, equal to the entropy difference
between $\tau_{\beta}^{(S)}$ and $\rho_S$.
\par
An interesting scenario occurs when $\Lambda$ maps the initial state into $\tau_{\beta}^{(S)}$.
In this case, we see that the work we obtain in the second transformation (the one involving the
whole thermal bath) is proportional to the sole mutual information, so that work is exclusively extracted
from the correlations between system and catalyst. The amount of work in this case (see also
Ref.~\cite{perarnau-llobet_extractable_2015}, Sec.~VI B) is
\begin{equation}
\Delta W_2^{\text{corr}} = \frac{1}{\beta} I \left( \tilde{S} : \tilde{A} \right) =
\frac{1}{\beta} \left( S(\tau_{\beta}^{(S)}) - S(\rho_S) \right),
\end{equation}
where the quantity is still non-negative, since $\Lambda$ can map $\rho_S$ into $\tau_{\beta}^{(S)}$
only if $S( \rho_S ) \leq S( \tau_{\beta}^{(S)} )$, see Sec.~\ref{asympt_work_extr}.
\section{Technical results}
\label{technical_tools}
In this section we show some of the technical results we have used to analyse the generic cycle on passive states.
\begin{technical_result} \label{sequence_solution}
Consider the sequence of real numbers $\left\{ x_j \right\}_{a}^{b}$, those elements are linked by the following
set of equations,
\begin{equation*}
x_j = \left( 1 + \lambda \right) x_{j+1} - \lambda \, x_{j+2} \quad ; \quad j = a, \ldots , b - 2 ,
\end{equation*}
where $\lambda \in \R$ and $a, b \in \N$, $a \leq b - 2$. Then, the elements of this sequence
can be expressed in terms of $x_{b-1}$ and $x_{b}$ as
\begin{equation*}
x_j = \T( b - ( j + 1 ) , \lambda ) \, x_{b-1} - \lambda \, \T( b - ( j + 2 ) , \lambda) \, x_{b} \quad ; \quad j = a, \ldots , b - 2,
\end{equation*}
where $\T(h, \lambda) = \sum_{l = 0}^{h} \lambda^l = \frac{1 - \lambda^{h+1}}{1 - \lambda}$.
\end{technical_result}
\begin{proof}
If we insert the solution into the set of equations, we find
\begin{align*}
\T( b - ( j + 1 ) , \lambda ) \, x_{b-1} - \lambda \, \T( b - ( j + 2 ) , \lambda) \, x_{b} &=
( 1 + \lambda ) \T( b - ( j + 2 ) , \lambda ) \, x_{b-1} - \lambda ( 1 + \lambda ) \, \T( b - ( j + 3 ) , \lambda) \, x_{b} \\
&- \lambda \, \T( b - ( j + 3 ) , \lambda ) \, x_{b-1} + \lambda^2 \, \T( b - ( j + 4 ) , \lambda) \, x_{b}
\end{align*}
for $j$ taking values from $a$ to $b-2$. We can re-organise the above equation, and we find that it is satisfied iff
\begin{subequations}
\begin{align}
\label{cond_1}
\T( b - ( j + 1) , \lambda) &= ( 1 + \lambda ) \, \T( b - ( j + 2 ) , \lambda) - \lambda \, \T( b - ( j + 3 ) , \lambda)
\quad ; \quad j = a, \ldots , b - 2, \\
\label{cond_2}
\T( 0 , \lambda) &= ( 1 + \lambda ) \, \T( -1 , \lambda) - \lambda \, \T( -2 , \lambda).
\end{align}
\end{subequations}
These two equalities easily follow from the definition of $\T(h, \lambda)$, as it can be check by replacing
this coefficient with its explicit form in both Eq.~\eqref{cond_1} and \eqref{cond_2}.
\end{proof}
\begin{technical_result} \label{positive_probabilities}
The probability distribution of the state $\rho_M$ is positive and normalised.
\end{technical_result}
\begin{proof}
Let us consider the probabilities $\q_j$ for $j = 0 , \ldots , m-1$, as given in Eq.~\eqref{machine_prob_1}.
If we replace $j$ with $j' = m - j$, then the main coefficient in the equation becomes
\begin{align*}
\T_1(j') - \frac{\p_2}{\p_1} \, D(m,n) \, \T_1 ( j' - 1 ) &=
\frac{ \T_1(j') \, \T_1(m-1) - \T_1(j'-1) \, \T_1(m) }
{ \T_1(m-1) + \frac{\p_1}{\p_2} \T_2(n-1) } \\
&+ \frac{\p_1}{\p_2} \frac{ \T_1(j') \, \T_2(n-1) - \T_1(j'-1) \, \T_2(n-2) }
{ \T_1(m-1) + \frac{\p_1}{\p_2} \T_2(n-1) }.
\end{align*}
It is clear that the denominator is positive, as $\T_1(h)$ and $\T_2(h)$ are positive for all $h \in \Z$.
We need to show that the nominator is positive as well. The nominator of the first term can be reduced to
\begin{equation*}
\T_1(j') \, \T_1(m-1) - \T_1(j'-1) \, \T_1(m) = \T_1(m-1) - \T_1(j'-1) = \sum_{l=j'}^{m-1} \left( \frac{\p_0}{\p_1} \right)^l \geq 0,
\end{equation*}
where the last equality follows from the fact that  $j' = 1 , \ldots , m$.
The nominator of the second term can be expressed as
\begin{equation*}
\T_1(j') \, \T_2(n-1) - \T_1(j'-1) \, \T_2(n-2) = \T_1(j'-1) \left( \frac{\p_1}{\p_2} \right)^{n-1}
+ \T_2(n-2) \left( \frac{\p_0}{\p_1} \right)^{j'}
+ \left( \frac{\p_0}{\p_1} \right)^{j'} \left( \frac{\p_1}{\p_2} \right)^{n-1} > 0.
\end{equation*}
Thus, the probabilities $\left\{ q_j \right\}_{j=0}^{m-1}$ are positive when $\q_{m+n-1}$ is positive.
\par
We can now focus on the probabilities $\q_j$ for $j = m , \ldots , m+n-3$, as given in Eq.~\eqref{machine_prob_2}.
By replacing $j$ with $j' = m + n - (j + 2)$ we obtain that the main coefficient in the equation becomes
\begin{align*}
\T_2(j') \, D(m,n) - \frac{\p_1}{\p_2}  \, \T_2 ( j' - 1 ) &=
 \left( \frac{\p_1}{\p_2} \right) \, \frac{ \T_2(j') \, \T_1(m) - \T_2(j'-1) \, \T_1(m-1) }
{ \T_1(m-1) + \frac{\p_1}{\p_2} \T_2(n-1) } \\
&+ \left( \frac{\p_1}{\p_2} \right)^2 \frac{ \T_2(j') \, \T_2(n-2) - \T_2(j'-1) \, \T_2(n-1) }
{ \T_1(m-1) + \frac{\p_1}{\p_2} \T_2(n-1) }.
\end{align*}
As before, the denominator is positive, as $\T_1(h)$ and $\T_2(h)$ are both positive $\forall \, h \in \Z$.
The nominator of the first term can be reduced to
\begin{equation*}
\T_2(j') \, \T_1(m) - \T_2(j'-1) \, \T_1(m-1) = \T_2(j'-1) \left( \frac{\p_0}{\p_1} \right)^{m}
+ \T_1(m-1) \left( \frac{\p_1}{\p_2} \right)^{j'}
+ \left( \frac{\p_1}{\p_2} \right)^{j'} \left( \frac{\p_0}{\p_1} \right)^{m} > 0.
\end{equation*}
The nominator of the second term can be expressed as
\begin{equation*}
\T_2(j') \, \T_2(n-2) - \T_2(j'-1) \, \T_2(n-1) = \T_2(n-2) - \T_2(j'-1) = \sum_{l=j'}^{n-2} \left( \frac{\p_1}{\p_2} \right)^l \geq 0,
\end{equation*}
where the last equality follows from the fact that  $j' = 1 , \ldots , n-2$.
Thus, the probabilities $\left\{ q_j \right\}_{j=m}^{m+n-3}$ are positive when $\q_{m+n-1} > 0$.
\par
In Eq.~\eqref{machine_prob_3}, we showed that $\q_{m+n-2}$ is related to $\q_{m+n-1} $ by the multiplicative
coefficient $D(m,n)$, which can be easily shown to be positive for any integer $m, n \geq 1$. Finally, the normalisation
condition force $\q_{m+n-1}  > 0$, and implies the probability distribution of $\rho_M$ to be positive and normalised.
\end{proof}

\end{document}